\documentclass[aps,prx,reprint,twocolumn,superscriptaddress,floatfix,nofootinbib,longbibliography]{revtex4-2}
\usepackage{graphicx,amsmath,amsfonts,amssymb,amsthm,xr}
\usepackage{epsfig,amsmath,amssymb,color,dsfont,upgreek,physics}
\usepackage{mathrsfs}
\usepackage{mathtools}
\usepackage{bbold}
\usepackage{float}
\usepackage{helvet}
\usepackage[caption=false]{subfig}

\usepackage[bookmarks=true,colorlinks,linkcolor=OrangeRed,urlcolor=NavyBlue,citecolor=RoyalBlue]{hyperref}
\usepackage[dvipsnames]{xcolor}
\usepackage{orcidlink}

\usepackage{graphicx}
\usepackage{dcolumn}
\usepackage{bm}
\usepackage{color}
\usetikzlibrary{quantikz2}

\definecolor{mygold}{rgb}{0.5,0.6,0.7}

\begin{document}

\title{String Breaking and Glueball Dynamics in $2+1$D Quantum Link Electrodynamics}

\author{Jiahao Cao}
\affiliation{Department of Physics and Arnold Sommerfeld Center for Theoretical Physics (ASC), Ludwig Maximilian University of Munich, 80333 Munich, Germany}
\affiliation{Munich Center for Quantum Science and Technology (MCQST), 80799 Munich, Germany}

\author{Rohan Joshi}
\affiliation{Max Planck Institute of Quantum Optics, 85748 Garching, Germany}
\affiliation{Department of Physics and Arnold Sommerfeld Center for Theoretical Physics (ASC), Ludwig Maximilian University of Munich, 80333 Munich, Germany}
\affiliation{Munich Center for Quantum Science and Technology (MCQST), 80799 Munich, Germany}

\author{Yizhuo Tian}
\affiliation{Department of Physics and Arnold Sommerfeld Center for Theoretical Physics (ASC), Ludwig Maximilian University of Munich, 80333 Munich, Germany}
\affiliation{Munich Center for Quantum Science and Technology (MCQST), 80799 Munich, Germany}

\author{N.~S.~Srivatsa${}^{\orcidlink{0000-0001-6433-450X}}$}
\email{srivatsa.prasanna@lmu.de}
\affiliation{Max Planck Institute of Quantum Optics, 85748 Garching, Germany}
\affiliation{Department of Physics and Arnold Sommerfeld Center for Theoretical Physics (ASC), Ludwig Maximilian University of Munich, 80333 Munich, Germany}
\affiliation{Munich Center for Quantum Science and Technology (MCQST), 80799 Munich, Germany}

\author{Jad C.~Halimeh${}^{\orcidlink{0000-0002-0659-7990}}$}
\email{jad.halimeh@lmu.de}
\affiliation{Department of Physics and Arnold Sommerfeld Center for Theoretical Physics (ASC), Ludwig Maximilian University of Munich, 80333 Munich, Germany}
\affiliation{Max Planck Institute of Quantum Optics, 85748 Garching, Germany}
\affiliation{Munich Center for Quantum Science and Technology (MCQST), 80799 Munich, Germany}
\affiliation{Department of Physics, College of Science, Kyung Hee University, Seoul 02447, Republic of Korea}

\date{\today}

\begin{abstract}
At the heart of quark confinement and hadronization, the physics of flux strings has recently become a focal point in the field of quantum simulation of high-energy physics (HEP). Despite considerable progress, a detailed understanding of the behavior of flux strings in quantum simulation-relevant lattice formulations of gauge theories has remained limited to the lowest truncations of the gauge field, which are severely limited in their ability to draw conclusions about the quantum field theory limit. Here, we employ tensor network simulations to investigate the behavior of flux strings in a quantum link formulation of $2+1$D quantum electrodynamics (QED) with a spin-$1$ representation of the gauge field. We first map out the ground-state phase diagram of this model in the presence of two spatially separated static charges, revealing distinct microscopic processes responsible for string breaking, including a two-stage breaking mechanism not possible in the spin-$\frac{1}{2}$ formulation. Starting in different initial product state string configurations, we then explore far-from-equilibrium quench dynamics across various parameter regimes, demonstrating genuine $2+1$D real-time string breaking and glueball-like bound state formation, with the latter not possible in the spin-$\frac{1}{2}$ formulation. In and out of equilibrium, we consider different values and placements of the static charges. Finally, we provide efficient qudit circuits for a quantum simulation experiment in which our results can be observed in state-of-the-art ion-trap setups. Our findings lay the groundwork for quantum simulations of flux strings towards the quantum field theory limit.
\end{abstract}

\maketitle
{\hypersetup{linkcolor=mygold}
\tableofcontents}
\section{Introduction}

\begin{figure*}
\includegraphics[width=0.95\linewidth]{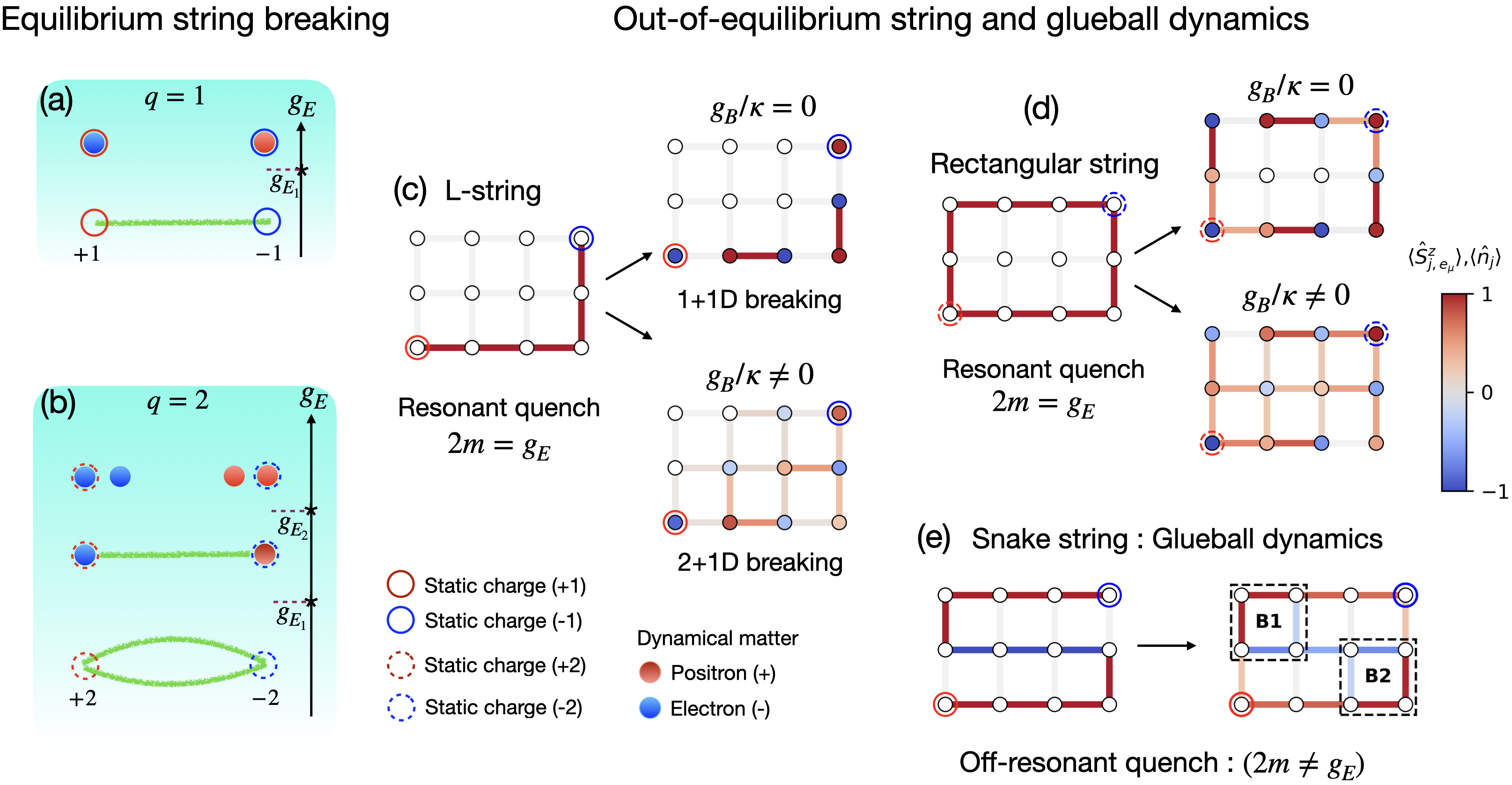}
\caption{Schematic illustrations and real-time snapshots of equilibrium and out-of-equilibrium string breaking and dynamics in the $2+1$D spin-$1$ $\mathrm{U}(1)$ QLM. Equilibrium string breaking: (a) Schematic showing the breaking of a string of electric flux (illustrated in green) connecting opposite static charges of magnitude $q=1$, where increasing the electric field strength $g_E$ beyond a critical value $g_{E_1}$ leads to the creation of a pair of dynamical matter particles comprising of a positron and an electron. (b) Schematic of two-stage string breaking between opposite static charges of magnitude $q=2$. The minimal possible string is a closed loop connecting static charges. At the first critical field value $g_{E_1}$, the closed string partially breaks into a single string, producing a positron-electron pair, followed by a complete string breaking at the second critical field value $g_{E_2}$ with an additional positron-electron pair. Out-of-equilibrium dynamics: (c) Real-time snapshots of the dynamical breaking of an initial L-string at resonance $2m=g_E$. In the absence of the plaquette term ($g_B/\kappa=0$), string breaking is restricted to the initial configuration, resulting in effectively $1+1$D dynamics. When $g_B/\kappa \neq 0$, genuine $2+1$D string breaking occurs beyond the initial minimal string. (d) Snapshots illustrating the resonant breaking of a rectangular string connecting opposite static charges of magnitude $q=2$. For $g_B = 0$, the string breaks only along its initial configuration, whereas for $g_B \neq 0$, breaking also occurs outside the initial string. (e) Snapshots illustrating dynamical glueball formation (B1,B2) from a snake initial string quenched off resonance ($2m\neq g_E$). }
\label{fig:intro_fig}
\end{figure*}

Quark confinement and string breaking are intimately connected phenomena underlying HEP \cite{Weinberg1995QuantumTheoryFields,Gattringer2009QuantumChromodynamicsLattice,Zee2003QuantumFieldTheory,Ellis2003QCDColliderPhysics}. Gauge theories---fundamental models constituting the Standard Model of particle physics and stipulating intrinsic relations between the distribution of charged matter and gauge fields---are natural venues to investigate both. In gauge theories, static charges are connected by an electric flux tube whose energy increases linearly with the separation between the charges, leading to confinement at intermediate distances. In the presence of dynamical fermions, however, this confining string is not indefinitely stable and once the energy stored in the flux tube exceeds the mass threshold for particle-antiparticle pair creation, the string breaks, resulting in the screening of the static charges by a newly created pair of dynamical charges; see Fig.~\ref{fig:intro_fig}(a,b). 

Lattice formulations of gauge theories, known as lattice gauge theories (LGTs) \cite{Kogut1975HamiltonianFormulationWilsons,Kogut1979AnIntroductionToLatticeGaugeTheory,Rothe2012LatticeGaugeTheories}, were devised for the purpose of studying quark confinement from first principles \cite{Wilson1974ConfinementQuarks,Wilson1977QuarksStringsLattice}. Indeed, confinement is a low-energy nonperturbative strong-coupling phenomenon, and continuum quantum field theory has no reliable way of addressing strong-coupling physics. LGTs, on the other hand, provide both the conceptual framework and the calculational tools to do so. Among these tools, Monte Carlo methods \cite{Creutz1979MonteCarloStudy,Creutz1980MonteCarloStudy,Creutz1983MonteCarloComputations,Creutz1988LatticeGaugeTheory,Creutz1989LatticeGaugeTheories,montvay1994quantum,Kieu1994MonteCarloSimulations,Hackett2019DigitizingGaugeFields} have been prominent and have led to a lot of successful results in studying equilibrium properties of string breaking in LGTs \cite{Philipsen1998StringBreakingNonAbelianGaugeTheories,Knechtli1998StringBreakingSU2GaugeTheory,Bali2005ObservationStringBreakingQCD,Pepe2009FromDecayToCompleteBreaking,Bulava2019StringBreakingLightStrangeQuarksQCD}. However, Monte Carlo techniques have been traditionally mostly limited to equilibrium and low-density regimes due to the sign problem \cite{Troyer2005ComputationalComplexityFundamental,deforcrand2010simulatingqcdfinitedensity}. 

A powerful framework that avoids the sign problem is tensor networks (TNs) \cite{Schollwock2011DensitymatrixRenormalizationGroup,Orus2014PracticalIntroductionTensorNetworks,Montangero2018IntroductionTensorNetwork,Orus2019TensorNetworksComplex,Paeckel2019TimeevolutionMethodsMatrixproduct}, which have allowed the simulation of quantum many-body models, including LGTs in one, two, and even three spatial dimensions \cite{Dalmonte2016LatticeGaugeTheory,Magnifico2024TensorNetworksLattice}. Recently, TN simulations in one and two spatial dimensions of LGT flux string behavior in and out of equilibrium have also been performed \cite{Kuhn2015NonAbelianStringBreaking,Pichler2016RealTimeDynamicsU1,Kasper2016SchwingerPairProduction,Buyens2017RealTimeSimulationSchwingerEffect,Kuno2017QuantumSimulation$1+1$dimensional,Sala2018VariationalStudyU1,Spitz2019SchwingerPairProduction,Park2019GlassyDynamicsFromQuarkConfinement,Surace2020LatticeGaugeTheories,Magnifico2020RealTimeDynamics,Notarnicola2020RealtimedynamicsQuantumSimulation,Chanda2020ConfinementLackThermalization,Borla2025StringBreaking$2+1$D,Xu2025StringBreakingDynamics,Barata2025RealTimeSimulationJetEnergyLoss,Marcantonio2025RougheningDynamicsElectric,Xu2025TensorNetworkStudyRoughening,artiaco2025outofequilibriumdynamicsu1lattice,cataldi2025realtimestringdynamics21d}. Despite their great power, TN methods are also limited in accessible evolution times due to rapid entanglement growth particularly in higher dimensions.

This has been a major motivation behind another powerful tool with which LGTs can be studied: Quantum simulation \cite{Bloch2012QuantumSimulationUltracoldQuantumGases,Georgescu2014QuantumSimulation,Gross2017QuantumSimulations,Altman2021QuantumSimulators,Alexeev2021QuantumComputerSystemsScientificDiscovery}. Recent years have witnessed a strong drive towards the quantum simulation of HEP \cite{Byrnes2006SimulatingLatticeGauge, Dalmonte2016LatticeGaugeTheory, Zohar2015QuantumSimulationsLattice, Aidelsburger2021ColdAtomsMeet, Zohar2021QuantumSimulationLattice, 
Barata2022MediumInducedJetBroadening,Klco2022StandardModelPhysics,Barata2023QuantumSimulationInMediumQCDJets,Barata2023RealTimeDynamicsofHyperonSpin, Bauer2023QuantumSimulationHighEnergy, Bauer2023QuantumSimulationFundamental,
DiMeglio2024QuantumComputingHighEnergy, Cheng2024EmergentGaugeTheory, Halimeh2022StabilizingGaugeTheories, Cohen2021QuantumAlgorithmsTransport,Barata2025ProbingCelestialEnergy, Lee2025QuantumComputingEnergy, Turro2024ClassicalQuantumComputing,Halimeh2023ColdatomQuantumSimulators,Bauer2025EfficientUseQuantum,Halimeh2025QuantumSimulationOutofequilibrium}. The stated overarching purpose is to be able to study HEP phenomena from first principles on quantum simulators that can provide snapshots of the real-time dynamics. This has the potential to create a viable venue complementary to dedicated particle colliders, where \textit{ab initio} studies of $3+1$D QCD can one day be undertaken. Even though the field is still far away from this holy grail, there has been a flurry of impressive quantum simulation experiments probing various LGT phenomena in one and two spatial dimensions \cite{Martinez2016RealtimeDynamicsLattice, Klco2018QuantumclassicalComputationSchwinger,Gorg2019RealizationDensitydependentPeierls, Schweizer2019FloquetApproachZ2, Mil2020ScalableRealizationLocal, Yang2020ObservationGaugeInvariance, Wang2022ObservationEmergent$mathbbZ_2$, Su2023ObservationManybodyScarring, Zhou2022ThermalizationDynamicsGauge, Wang2023InterrelatedThermalizationQuantum, Zhang2025ObservationMicroscopicConfinement, Zhu2024ProbingFalseVacuum, Ciavarella2021TrailheadQuantumSimulation, Ciavarella2022PreparationSU3Lattice, Ciavarella2023QuantumSimulationLattice-1, Ciavarella2024QuantumSimulationSU3, 
Gustafson2024PrimitiveQuantumGates, Gustafson2024PrimitiveQuantumGates-1, Lamm2024BlockEncodingsDiscrete, Farrell2023PreparationsQuantumSimulations-1, Farrell2023PreparationsQuantumSimulations, 
Farrell2024ScalableCircuitsPreparing,
Farrell2024QuantumSimulationsHadron, Li2024SequencyHierarchyTruncation, Zemlevskiy2025ScalableQuantumSimulations, Lewis2019QubitModelU1, Atas2021SU2HadronsQuantum, ARahman2022SelfmitigatingTrotterCircuits, Atas2023SimulatingOnedimensionalQuantum, Mendicelli2023RealTimeEvolution, Kavaki2024SquarePlaquettesTriamond, Than2024PhaseDiagramQuantum, Angelides2025FirstorderPhaseTransition, Gyawali2025ObservationDisorderfreeLocalization,  
Mildenberger2025Confinement$$mathbbZ_2$$Lattice, Schuhmacher2025ObservationHadronScattering, Davoudi2025QuantumComputationHadron, Saner2025RealTimeObservationAharonovBohm, Xiang2025RealtimeScatteringFreezeout, Wang2025ObservationInelasticMeson,li2025frameworkquantumsimulationsenergyloss,mark2025observationballisticplasmamemory,froland2025simulatingfullygaugefixedsu2,Hudomal2025ErgodicityBreakingMeetsCriticality,hayata2026onsetthermalizationqdeformedsu2} including flux strings \cite{Cochran2025VisualizingDynamicsCharges, Gonzalez-Cuadra2025ObservationStringBreaking, Crippa2024AnalysisConfinementString, De2024ObservationStringbreakingDynamics, Liu2024StringBreakingMechanism, Alexandrou2025RealizingStringBreaking,Cobos2025RealTimeDynamics2+1D}. As impressive as these experiments are, they are mostly restricted to a two-level discretization of the gauge field, and those on a $2$d lattice have mostly effectively probed $1+1$D string dynamics due to the absence of a plaquette term \cite{Tian2025RolePlaquetteTerm}.

Of particular interest in the quantum simulation of string dynamics are U$(1)$ quantum link model (QLM) formulations \cite{Chandrasekharan1997QuantumLinkModels,Wiese2013UltracoldQuantumGases,Kasper2017ImplementingQuantumElectrodynamics} of QED, where the gauge and electric fields are represented by spin-$S$ raising and $z$ operators, respectively. These constitute some of the earliest works on string breaking in LGTs \cite{Banerjee2012AtomicQuantumSimulation} and their connection to QED in $1+1$D has been established in this context \cite{Hebenstreit2013RealTimeDynamicsStringBreaking}. In the Kogut--Susskind limit $S\to\infty$, U$(1)$ QLMs retrieve the low-energy limit of lattice QED. However, it has been shown that even at small values of $S$, many features of lattice QED can be achieved in and out of equilibrium \cite{Buyens2017FiniteRepresentationLatticeGaugeTheories,Zache2022TowardContinuumLimit,Halimeh2022achievingquantum}.

A scalable \texttt{QuEra} analog Rydberg-atom quantum simulation experiment studied string breaking in the U$(1)$ QLM for $S=\frac{1}{2}$ in and out of equilibrium on a hexagonal lattice without a plaquette term \cite{Gonzalez-Cuadra2025ObservationStringBreaking}. The absence of a plaquette term limits the observed string phenomenology to $1+1$D and gives rise to string breaking dynamics resembling Rabi oscillations \cite{Tian2025RolePlaquetteTerm}. The spin-$\frac{1}{2}$ formulation renders the gauge coupling term, which encodes the energy of the electric field, an inconsequential energetic constant, further limiting its connection to lattice QED where a tunable gauge coupling term is essential for confinement. In the \texttt{QuEra} experiment, confinement is induced by artificially adding an explicit term proportional to the electric flux at each link. Furthermore, the spin-$\frac{1}{2}$ formulation restricts strings to minimal Manhattan-distance configurations for reasonable choice of static charges, which prevents the necessary conditions for glueball-like closed loops to dynamically emerge \cite{Xu2025StringBreakingDynamics,cataldi2025realtimestringdynamics21d}. These limitations are fundamental and stand in the way of utilizing the full power of QLMs to infer the physics of the Kogut--Susskind limit. This motivates considering a $2+1$D U$(1)$ QLM with a higher-level representation of the gauge field and thoroughly investigating its flux-string behavior in and out of equilibrium as a guide to future quantum simulation experiments. Another motivation for this undertaking is that currently TN simulations are still able to outperform the aforementioned string-breaking experiments in size and maximally accessible evolution time. It is expected that with higher-level representations of the gauge field, which drastically increase the dimension of the Hilbert space, TN simulations will face a greater challenge due to a more rapid increase in entanglement with evolution time relative to the spin-$\frac{1}{2}$ case. This therefore provides a possible venue of quantum advantage particularly for qudit processors that are ideally suited to implement larger-$S$ representations of the gauge fields.

Qudits \cite{Ringbauer_2022} have recently emerged as a powerful platform, with numerous theoretical proposals \cite{ciavarella2022conceptualaspectsoperatordesign,Popov2024variational,Calajo2024digital,kürkçüoglu2024quditgatedecompositiondependence,ballini2025symmetryverificationnoisyquantum,gaz2025quantumsimulationnonabelianlattice,jiang2025nonabeliandynamicscubeimproving,PhysRevD.110.014507, joshi2025probinghadronscatteringlattice, joshi2025efficientquditcircuitquench}, and experimental demonstrations of single-plaquette dynamics in higher-spin formulations \cite{Meth_Zhang_Haase_Edmunds_Postler_Jena_Steiner_Dellantonio_Blatt_Zoller_et} indicating that they offer a more natural and efficient alternative to qubits for the quantum simulation of LGTs by enabling the encoding of information in $d \geq 2$ levels. As we discuss in this work, qudits are particularly suited for simulating $2+1$D U$(1)$ QLM with higher spin formulations \cite{joshi2025efficientquditcircuitquench} as they preserve the native interaction range of the Hamiltonian terms.

In this work, we study flux strings in and out of equilibrium for the case of a $2+1$D U$(1)$ QLM with a spin-$1$ formulation where these limitations are immediately lifted. Using large-scale TN simulations, we calculate the ground-state phase diagram of this model in the presence of two spatially separated static charges in the confined regime. In particular, we study the effect of the placement and values of these static charges, showing how this can lead to rich physics, including a two-stage string-breaking process. We then calculate the quench dynamics starting in simple product state string configurations between these two static charges. The resulting dynamics is genuinely $2+1$D only in the presence of a finite plaquette term; see Fig.~\ref{fig:intro_fig}(c,d). Starting in so-called snake string configurations, Abelian analogues of glueballs dynamically emerge at high probability; see Fig.~\ref{fig:intro_fig}(e). Since this model is within reach of state-of-the-art digital qudit quantum simulators, we propose efficient qudit circuits that would enable the observation of our findings on quantum hardware in the near future. Our findings open the door to probing string breaking towards the quantum field theory limit, as they directly address the question of how the field truncation affects string behavior in and out of equilibrium.

\section{Model}

We consider a $2+1$D U$(1)$ QLM~\cite{Chandrasekharan1997QuantumLinkModels,Wiese2013UltracoldQuantumGases,Kasper2017ImplementingQuantumElectrodynamics} defined on a square lattice with Hamiltonian
\begin{align}\nonumber
\hat{H}=&\underbrace{-\kappa\sum_{\mathbf{j},\mu}\Big(s_{\mathbf{j},\mathbf{e}_{\mu}}
\hat{\phi}^{\dagger}_{\mathbf{j}}
\hat{U}_{\mathbf{j},\mathbf{e}_{\mu}}
\hat{\phi}_{\mathbf{j}+\mathbf{e}_{\mu}}
+\textrm{H.c.}\Big)}_{\hat{H}_\kappa}
+m\sum_{\mathbf{j}}s_{\mathbf{j}}\hat{\phi}^{\dagger}_{\mathbf{j}}\hat{\phi}_{\mathbf{j}}\\
&+g_{E}\sum_{\mathbf{j},\mu}\big(\hat{E}_{\mathbf{j},\mathbf{e}_{\mu}}\big)^2
-\underbrace{g_{B}\sum_{\Box}\Big(\hat{U}_{\Box}+\hat{U}^{\dagger}_{\Box}\Big)}_{\hat{H}_{\Box}}.
\label{eq:hamiltonian}
\end{align}
The first term, $\hat{H}_\kappa$, describes the minimal coupling with strength $\kappa$ between the matter fields $\hat{\phi}_{\mathbf{j}}$ located at lattice sites $\mathbf{j}=(j_x,j_y)^\intercal$ and representing hard-core bosons, whose canonical (anti)commutation relations are $\big\{\hat \phi_\textbf{j}, \hat \phi_{\textbf{j}'}^\dagger\big\} = \delta_{\textbf{j},\textbf{j}'}$, $\big[\hat \phi_\textbf{j}, \hat \phi_{\textbf{j}'}^\dagger\big] = \delta_{\textbf{j},\textbf{j}'}\big(1-2\hat\phi_\textbf{j}^\dagger\hat\phi_\textbf{j}\big)$, and $\big\{\hat \phi_\textbf{j}, \hat \phi_{\textbf{j}'}\big\}=\big[\hat \phi_\textbf{j}, \hat \phi_{\textbf{j}'}\big]=0$, and the gauge fields
$\hat{U}_{\mathbf{j},\mathbf{e}_\mu}$ residing on the links connecting sites
$\mathbf{j}$ and $\mathbf{j}+\mathbf{e}_\mu$, with $\mathbf{e}_\mu$ a unit vector in either the $x$ or $y$ spatial direction. To truncate the Hilbert space of the gauge and electric fields, we employ a spin-$1$ representation, which
constitutes the simplest nontrivial realization compatible with current quantum simulation platforms~\cite{Crippa2024AnalysisConfinementString}. Within this representation, the gauge and electric field operators are identified as the spin-$1$ operators
$\hat{U}_{\mathbf{j},\mathbf{e}_\mu}=\hat{S}^+_{\mathbf{j},\mathbf{e}_\mu}$ and $\hat{E}_{\mathbf{j},\mathbf{e}_{\mu}}=\hat{S}^z_{\mathbf{j},\mathbf{e}_\mu}$. The second term introduces a staggered mass $m$ for the matter fields. This formulation follows the Kogut--Susskind formulation for staggered fermions which results in staggered hopping and mass terms~\cite{Kogut1979AnIntroductionToLatticeGaugeTheory}. The staggering in the hopping is direction dependent, with $s_{\mathbf{j},\mathbf{e}_x}=+1$ and $s_{\mathbf{j},\mathbf{e}_y}=(-1)^{j_x}$. The mass staggering is
defined by $s_{\mathbf{j}}=(-1)^{j_x+j_y}$, such that the presence of a particle on an even site ($s_{\mathbf{j}}=+1$) corresponds to a positron with positive charge, while the absence of a particle on an odd site ($s_{\mathbf{j}}=-1$) represents an electron with negative charge. The third term accounts for the electric-field energy with coupling strength $g_E$. The final term, $\hat{H}_{\Box}$, represents the magnetic energy through a plaquette interaction with strength $g_B$ and the plaquette operator is defined as
\begin{align}
\hat{U}_\Box =\hat{U}_{\mathbf{j},\mathbf{e}_x}\hat{U}_{\mathbf{j}+\mathbf{e}_x,\mathbf{e}_y}
\hat{U}^\dagger_{\mathbf{j}+\mathbf{e}_y,\mathbf{e}_x}
\hat{U}^\dagger_{\mathbf{j},\mathbf{e}_y}.
\end{align}
Throughout this work we set $\kappa=1$. 

The Hamiltonian remains invariant under local gauge transformations generated by
\begin{equation}
\hat{G}_{\mathbf{j}} =
\hat{\phi}^{\dagger}_{\mathbf{j}} \hat{\phi}_{\mathbf{j}}
-\frac{1-(-1)^{\mathbf{j}}}{2}
-\sum_{\mu}\Big(
\hat{E}_{\mathbf{j},\mathbf{e}_{\mu}}
-
\hat{E}_{\mathbf{j}-\mathbf{e}_{\mu},\mathbf{e}_{\mu}}
\Big),
\end{equation}
which enforces Gauss's law constraint at each lattice site. Gauge-invariant states $\ket{\Psi}$ are simultaneous eigenstates of the generators $\hat{G}_{\mathbf{j}}$ at all $\mathbf{j}$: $\hat{G}_{\mathbf{j}}\ket{\Psi}=g_{\mathbf{j}}\ket{\Psi},\,\forall\mathbf{j}$, where the eigenvalues $g_\mathbf{j}$ are so-called background charges. Here, we set $g_\mathbf{j}=0$ at all sites $\mathbf{j}$ except for two: one even site $\mathbf{j}_\mathrm{e}$ and one odd site $\mathbf{j}_\mathrm{o}$ where we place the static charges, i.e., $g_{\mathbf{j}_\mathrm{e}}=-q$ and $g_{\mathbf{j}_\mathrm{o}}=+q$.

\section{Ground-state string breaking}
To investigate equilibrium string breaking, we consider a parameter regime deep in the confined phase, $m,g_E \gg \kappa$, where matter fluctuations are suppressed. In this regime, the interaction potential between static charges initially increases linearly with their separation, reflecting the formation of a string of electric flux connecting the charges. This linear confinement, however, does not persist indefinitely: when the energy stored in the flux string exceeds the threshold for positron-electron pair creation, the string breaks due to the emergence of dynamical charges (positron-electron pair) from the vacuum. 

We investigate this phenomenon using the density matrix renormalization group (DMRG) \cite{White1992DensityMatrixFormulationQuantumRenormalizationGroups,White1993DensityMatrixAlgorithmQuantumRenormalizationGroups,Schollwoeck2005DensityMatrixRenormalizationGroup} algorithm implemented in the \texttt{TenPy} library~\cite{tenpy} and convergence is ensured by employing matrix product states (MPS) \cite{Schollwock2011DensitymatrixRenormalizationGroup} with bond dimensions up to $\chi = 600$; see Appendix~\ref{app:convergence} for details. Ground states are computed while enforcing the presence of static charges through large energy penalty terms proportional to $(\hat{G}_{\mathbf{j}_{\mathrm{e}}}+q)^2$ and $(\hat{G}_{\mathbf{j}_{\mathrm{o}}}-q)^2$ \cite{Halimeh2020ReliabilityOfLatticeGaugeTheories} at the even ($\mathbf{j}_{\mathrm{e}}$) and odd ($\mathbf{j}_{\mathrm{o}}$) lattice sites, respectively, hosting static charges of magnitude $q$. At all other sites $\mathbf{j}$, a penalty term proportional to $\hat{G}_{\mathbf{j}}^2$ is added to project onto the sector of Gauss's law without charges.

Throughout this work, we consider configurations in which the positive (negative) static charge is placed on an odd (even) lattice site. This choice is consistent with the staggered fermion formulation, which allows dynamical matter with opposite charges (electron and positron) to be created on the same lattice sites as the imposed static charges, thereby enabling charge screening. The complementary scenario, in which this condition is violated, is discussed in Appendix \ref{alternate}. 

\begin{figure}
\centering
\includegraphics[width=\linewidth]{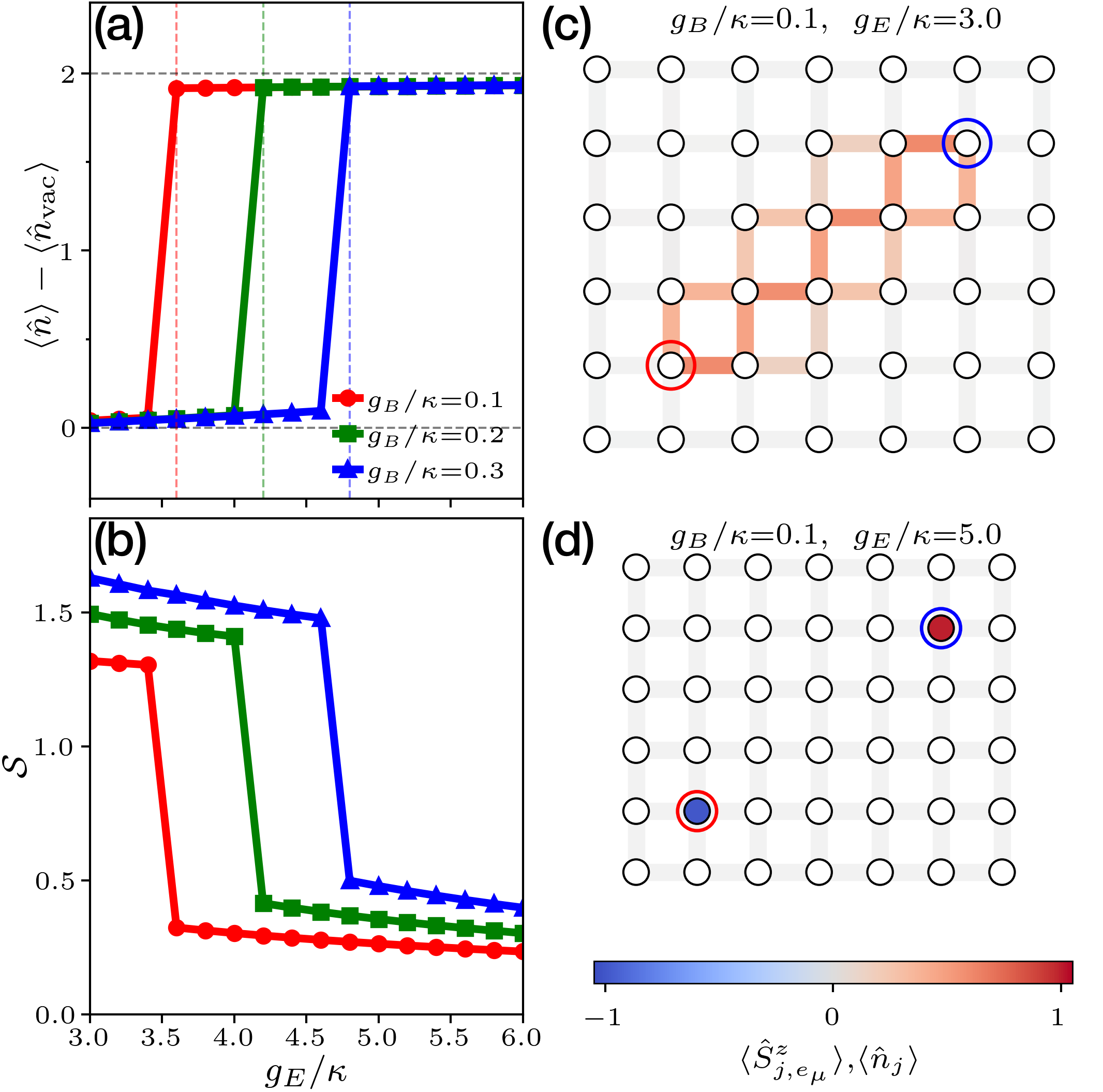}
\caption{Equilibrium string breaking between static charges of magnitude $q=1$ with $+q$ placed at the odd lattice site $(6,1)$ and $-q$ placed at the even lattice site $(10,4)$ on a rectangular lattice of dimension $L_x=16$ and $L_y=6$ for the $2+1$D spin-$1$ U$(1)$ QLM as the electric coupling $g_E$ is increased at fixed $g_B$. String breaking is signaled by (a) a sharp increase in the vacuum-subtracted matter density $\langle \hat{n} \rangle - \langle \hat{n}_{\mathrm{vac}} \rangle$, corresponding to the creation of a positron-electron pair, and (b) a rapid drop in the entanglement entropy $\mathcal{S}$. A finite magnetic coupling $g_B$ stabilizes the confining string, shifting the breaking point to larger values of $g_E$. (c) Representative string state (showing only the central region containing the $\pm q$ static charges, indicated by red and blue circles, respectively) b) prior to breaking, computed at $g_E/\kappa=3.0$ and $g_B/\kappa = 0.1$, where the two static charges are connected by a superposition of electric flux string configurations. (d) Broken-string state at $g_E/\kappa = 5.0$ and $g_B/\kappa = 0.1$, in which the string disappears and an electron is created at the site $(6,1)$ and positron at the site $(10,4)$ ensuring full screening of the static charges. For all cases, we use $m = 6$.}
\label{fig:combined_charge1}
\end{figure}

\begin{figure}
\centering
\includegraphics[width=1.0 \linewidth]{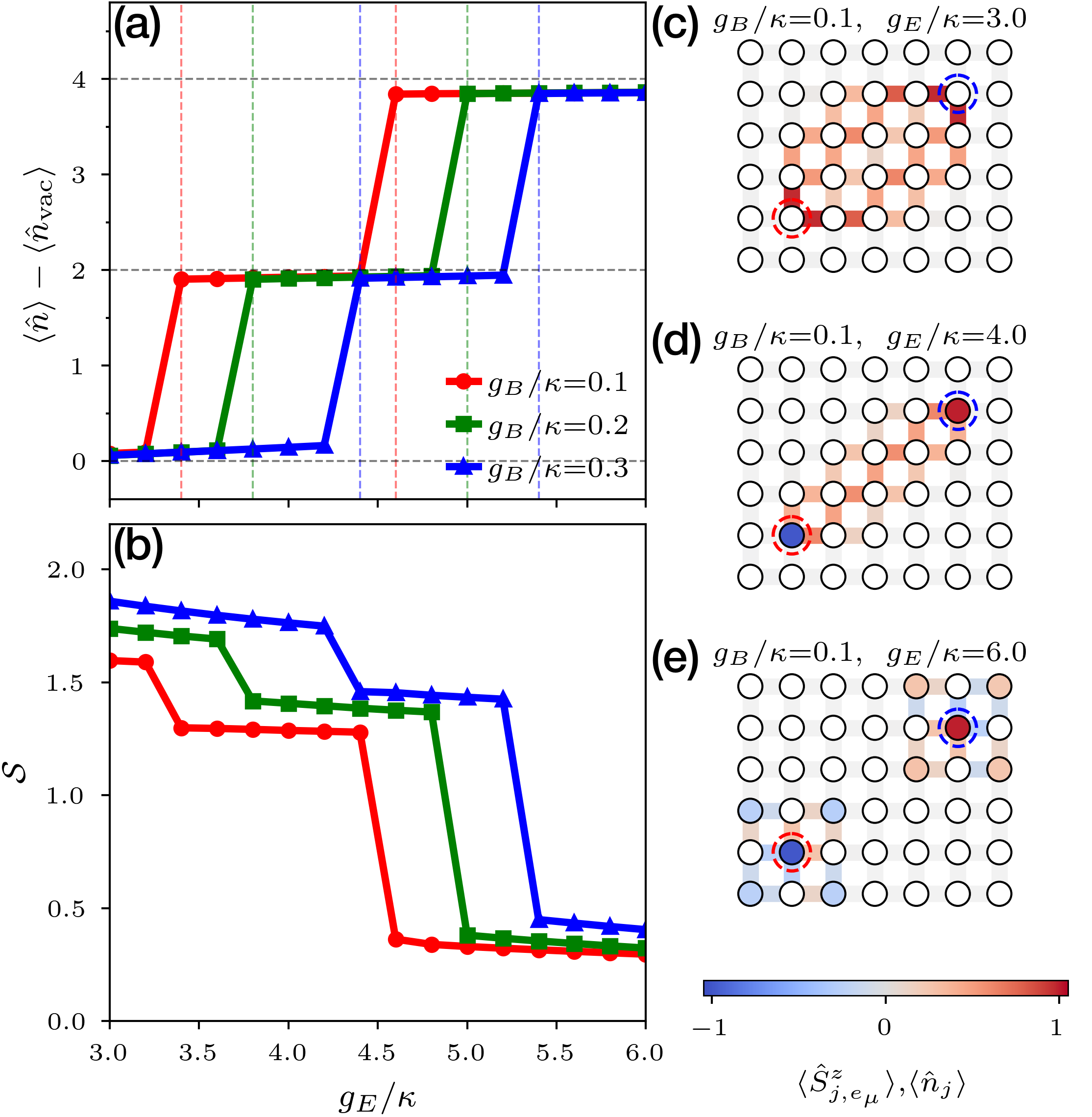}
\caption{Equilibrium string breaking between static charges of magnitude $q=2$ with $+q$ placed at the odd lattice site $(6,1)$ and $-q$ placed at the even lattice site $(10,4)$ on a rectangular lattice of dimension $L_x=16$ and $L_y=6$ for the $2+1$D spin-$1$ U$(1)$ QLM as the electric coupling $g_E$ is increased at fixed $g_B$. The two-stage string breaking process is reflected in (a) two successive jumps in the vacuum-subtracted matter density $\langle \hat{n} \rangle - \langle \hat{n}_{\mathrm{vac}} \rangle$, each corresponding to the creation of a positron-electron pair, and (b) two corresponding jumps in the entanglement entropy $\mathcal{S}$. A finite magnetic coupling $g_B$ stabilizes the confining strings, shifting both breaking points to larger values of $g_E$. (c) Representative string state (showing only the central patch containing static charges) prior to breaking, computed at $g_E/\kappa = 3.0$ and $g_B/\kappa  = 0.1$, where the static charges are connected by a superposition of electric flux string configurations. (d) Intermediate state after the first string breaking event, in which an electron is created at the site (6,1) and a positron at the site (10,4), while these two sites remain connected by a residual string. (e) Final fully broken string state after the second breaking event, in which the remaining string vanishes completely. The additional electron is distributed over the four odd lattice sites surrounding the lattice site $(6,1)$, while the additional positron is distributed over the four even lattice sites surrounding the lattice site $(10,4)$. All results are obtained for mass $m = 6$.}
\label{fig:combined_charge2}
\end{figure}

For gauge fields in the $S=1$ representation, Gauss's law restricts the allowed static charges to $q=1,2$, which are the minimal values capable of generating nontrivial string configurations as we discuss below. As a proxy for increasing the separation between static charges on a finite lattice, we tune the electric field coupling $g_E$ to effectively mimic charge separation. We consider a square lattice of size $L_x = 16$ and $L_y = 6$, with open (periodic) boundary conditions along the $x$ ($y$) direction, with two static charges placed at opposite corners of a rectangular patch of dimensions $l_1 \times l_2$, where $l_1 = 5$ and $l_2 = 4$ chosen in the center of the lattice to avoid boundary effects; see Fig.~\ref{fig:combined_charge1}.

To detect the onset of string breaking, we monitor the total matter density $\langle \hat{n} \rangle$ within a rectangular region that encloses the patch containing the static charges, supplemented by an additional surrounding layer of lattice sites to include any dynamical matter creation outside the string patch. This quantity is defined as
\begin{equation}
\langle \hat{n} \rangle = \sum_{\mathbf{j} \in \mathrm{patch}} (-1)^{\mathbf{j}} \, \langle \hat{n}_{\mathbf{j}} \rangle ,
\label{eq:matter}
\end{equation}
where $\hat{n}_{\mathbf{j}} = \hat{\phi}^{\dagger}_{\mathbf{j}} \hat{\phi}_{\mathbf{j}} - \big[1 - (-1)^{\mathbf{j}}\big]/2$ is the staggered fermion number operator. To isolate contributions arising from string breaking, we subtract the vacuum background by defining the vacuum subtracted matter density $\langle \hat{n} \rangle - \langle \hat{n}_{\mathrm{vac}} \rangle$, where $\langle \hat{n}_{\mathrm{vac}} \rangle$ denotes the corresponding expectation value in the absence of static charges. In addition, we compute the von Neumann entanglement entropy $\mathcal{S}$ associated with a bipartition of the system along the $y$ direction. We discuss different microscopic string breaking scenarios in the following. Videos of this physics are also included \cite{videos}.

\subsection{Static charges $\pm q$ with $q=1$}

As a first case, we consider static charges of magnitude $q=1$ placed at opposite corners of the rectangular patch described above. As shown in Fig.~\ref{fig:combined_charge1}(a,c), at low values of the electric field strength $g_E$, we observe a well-defined string configuration connecting the two static charges. This string is not a single classical configuration but rather a coherent linear superposition of multiple electric flux string configurations as is evident from the snapshot in Fig.~\ref{fig:combined_charge1}(c) showing the ground state configuration chosen at $g_E=3$ and $g_B=0.1$. We see that the largest-weight configuration is the one along the diagonal. This agrees with the findings of the $\texttt{QuEra}$ experiment \cite{Gonzalez-Cuadra2025ObservationStringBreaking}.

As $g_E$ is increased, the string becomes energetically unstable and undergoes a breaking transition clearly signaled by a pronounced jump in the matter density $\langle \hat{n} \rangle - \langle \hat{n}_{\mathrm{vac}} \rangle$ within the patch, as defined in Eq.~\eqref{eq:matter}. The magnitude of this jump is precisely two, consistent with the creation of dynamical matter that screens the two static charges of unit magnitude. The onset of string breaking is further accompanied by a sharp decrease in the von Neumann entanglement entropy, as shown in Fig.~\ref{fig:combined_charge1}(b), indicating a qualitative change in the structure of the ground state. This behavior is consistent with a transition from a highly entangled state in a superposition of multiple string configurations (see Fig.~\ref{fig:combined_charge1}(c)) to a broken-string state characterized by the presence of dynamical matter and a significantly reduced degree of entanglement, approaching that of a product state. As illustrated in the snapshot Fig.~\ref{fig:combined_charge1}(d), screening of static charges is clearly observed: a positron is produced at the even lattice site where a negative static charge is imposed, while an electron is generated at the odd lattice site where a positive static charge is imposed, in accordance with the staggered fermion convention discussed before.

We next investigate the role of magnetic fluctuations by tuning the magnetic field coupling $g_B$ while repeating the same parameter scan. We find that increasing $g_B$ enhances the stability of the string configuration, thereby shifting the string-breaking transition to larger values of $g_E$. This suggests that enhanced magnetic interactions stabilize the confining string by lowering its energy through string resonances and suppressing pair creation. This is consistent with what was observed in the $2+1$D $\mathbb{Z}_2$ LGT \cite{Borla2025StringBreaking$2+1$D}.

\subsection{Static charges $\pm q$ with $q=2$}

We now turn to a second, qualitatively distinct, scenario in which static charges of magnitude $q=2$ are enforced at the same lattice coordinates considered in the $q=1$ case. In this situation, the resulting string configuration occupies a significantly larger region of the rectangular patch at small values of $g_E$; see Fig.~\ref{fig:combined_charge2}(a,c). This behavior is consistent with Gauss's law, which dictates that the minimal flux configuration associated with higher static charge necessarily involves closed electric flux loops. This extended string profile is clearly visible in the ground-state snapshot shown in Fig.~\ref{fig:combined_charge2}(c).

Upon increasing the electric field strength $g_E$, we observe two distinct string-breaking transitions, in contrast to the single transition found for $q=1$. At the first transition, the string does not break completely; instead, it partially fragments via the creation of a single pair of dynamical matter—a positron at the even lattice site where a negative static charge is imposed, and an electron at the odd lattice site where a positive static charge is imposed. This intermediate configuration is illustrated in the ground-state snapshot of Fig.~\ref{fig:combined_charge2}(d), taken at a value of $g_E$ on the plateau immediately following the first transition. Correspondingly, the total vacuum subtracted matter density $\langle \hat{n} \rangle - \langle \hat{n}_{\mathrm{vac}} \rangle$ exhibits a discrete jump of two, consistent with the partial screening of static charges by a single generated positron-electron pair.

As $g_E$ is increased further, a second transition occurs in which the string fully breaks through the creation of an additional pair of dynamical matter, now around the sites hosting the static charges, as shown in Fig.~\ref{fig:combined_charge2}(e). This second breaking event is again accompanied by a further jump in $\langle \hat{n} \rangle - \langle \hat{n}_{\mathrm{vac}} \rangle$, signaling complete screening of the static charges. We note that the second pair of dynamical matter is distributed over the four matter sites adjacent to the static-charge sites. This behavior can be understood as follows: the odd lattice site is already occupied by an electron, and therefore any additional electron must reside on the four nearest neighboring odd lattice sites. Similarly, the additional positron can only be accommodated on the four nearest neighboring even lattice sites surrounding the even static-charge site that already hosts the positron generated during the first string-breaking event. As such, it is equally likely to find a dynamical-matter particle on either of the four matter sites surrounding a given static-charge site.

At both transition points, we observe a sharp drop in the von Neumann entanglement entropy, shown in Fig.~\ref{fig:combined_charge2}(b). In particular, the entanglement shows a decreasing trend across successive transitions, reflecting the reduction in the size of the coherent superposition of string configurations in the intermediate partially broken state and the emergence of a nearly product-state structure in the fully broken regime. A similar two-stage string-breaking mechanism has also been observed in $\mathrm{SU}(2)$ Yang--Mills theory through Monte Carlo techniques, involving an intermediate decay into lower-representation strings prior to complete string breaking \cite{Pepe2009FromDecayToCompleteBreaking}. It is important to note here that this two-stage string-breaking mechanism is not possible in the spin-$\frac{1}{2}$ formulation.

Finally, as in the $q=1$ case, we find that the magnetic interaction $g_B$ plays a stabilizing role for the string configurations. Increasing $g_B$ shifts both breaking transitions to larger values of $g_E$, indicating that magnetic fluctuations enhance the robustness against pair creation.

\section{Out-of-equilibrium string dynamics}
We now turn our attention to the out-of-equilibrium dynamics of string configurations. Our focus remains deep in the confined phase, where matter fluctuations are strongly suppressed. We prepare a variety of initial string configurations connecting static charges, with magnitude $q=1,2$ constrained by Gauss's law, and subsequently quench the system using the $2+1$D QLM Hamiltonian~\eqref{eq:hamiltonian} across different parameter regimes. The initial configuration is prepared by placing negative and positive static charges on an even (\(\mathbf{j}_\text{e}\)) and an odd (\(\mathbf{j}_\text{o}\)) site, respectively, and connecting them by a string of oriented electric field lines as permitted by Gauss's law. The sites containing the charges of magnitude $q$ satisfy the local constraints $\hat{G}_{\mathbf{j}_{\textrm{o/e}}}\ket{\Psi} = \pm q\ket{\Psi}$, while at all other sites $\hat{G}_{\mathbf{j}}|\Psi\rangle = 0$. 

To investigate the dynamics of the string and its subsequent breaking, we monitor three key quantities. The first is the fidelity, defined as the overlap of the time-evolved wave function $\ket{\psi}(t)=e^{-\imath \hat{H}t}\ket{\psi_0}$ with respect to the initial string state $\ket{\psi_0}$,
\begin{align}
\mathcal{F}(t) = \lvert \braket{\psi_0}{\psi(t)} \rvert^2 ,
\end{align}
which measures how much of the original state persists at time \(t\).
Second, we track the overlap
\begin{align}
\mathcal{P}_{\gamma \neq \gamma_{\text{i}}}
= \sum_{\gamma \neq \gamma_{\text{i}}}
\lvert \braket{\psi_\gamma}{\psi(t)} \rvert^2 ,
\end{align}
with all other string configurations \(\gamma\) other than the initial one \(\gamma_{\text{i}}\) but with the same length, providing a measure of transitions between different string configurations that also provides a direct measure of how $2+1$D the dynamics is. Lastly, we also track the total matter occupation $\langle \hat{n} \rangle$ within the patch defined in Eq.~\eqref{eq:matter}.

The dynamics of strings is simulated using the time-dependent variational principle (TDVP) algorithm \cite{Haegeman2011TimeDependentVariationalPrinciple,Bauernfeind2020TimeDependentVariational,Yang2020TimedependentVariationalPrinciple} as implemented in the \texttt{Matrix Product Toolkit}~\cite{mptoolkit}. We consider string configurations defined within patches of sizes $\ell_1=4$ and $\ell_2=3$ embedded in a lattice of dimensions $L_x=6$ and $L_y=6$, with open boundary conditions. To ensure numerical convergence, we employ bond dimensions up to $\chi=500$ and use TDVP time step of $\delta t=0.01$; see Appendix~\ref{app:convergence} for details. Videos of the dynamics are also included \cite{videos}.

\subsection{Resonant string-breaking dynamics}

Out of equilibrium, string breaking occurs when the energy accumulated in an electric flux string becomes comparable to the cost of producing a dynamical positron-electron pair. In the spin-$1$ QLM, for instance, when the electric-field eigenvalue on a link is $S^z_{\mathbf{j},\mathbf{e}_\mu}=+1$, each link in the string contributes an energy $g_E$. Consequently, a string segment consisting of $n$ such consecutive links carries a total energy $n g_E$. The breaking of such a segment occurs via an $n^\mathrm{th}$-order process, in which all $n$ links are flipped to $S^z_{\mathbf{j},\mathbf{e}_\mu}=0$, accompanied by the creation of a positron-electron pair at the two endpoints of the link costing an energy of $2m$. This mechanism leads to the resonance condition $2m = n g_E$. Within the staggered fermion formulation, only string segments containing an odd number of links can participate in this process, implying that $n$ is necessarily odd. In this work, we quench various initial string configurations using the QLM Hamiltonian, tuning the parameters to the first-order resonance condition $2m = g_E$; see Fig.~\ref{fig:intro_fig}(c,d).

\begin{figure}
\includegraphics[width=1.0 \linewidth]{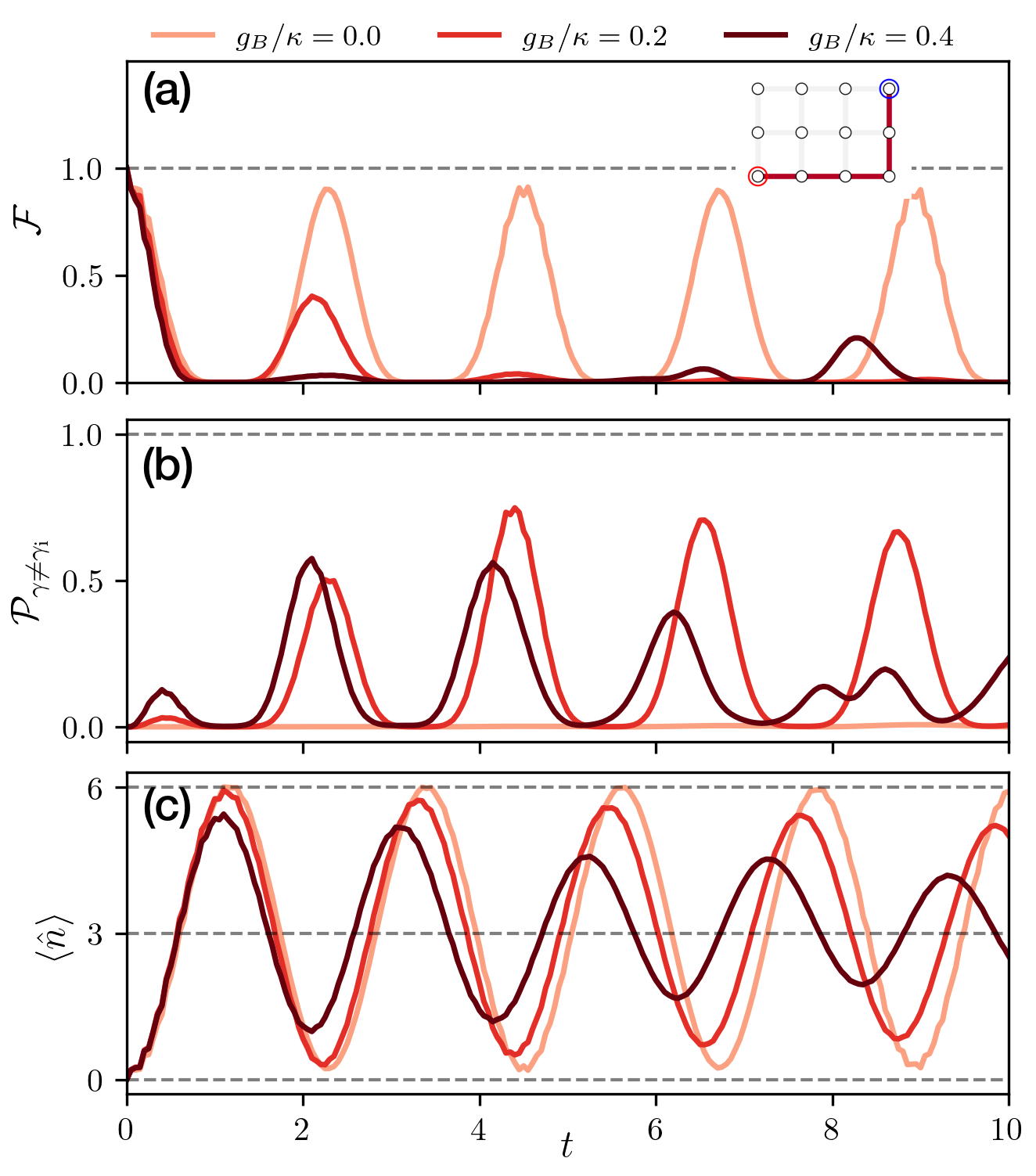}
\caption{Resonant dynamics of an L-string (see inset) between static charges of magnitude $q=1$ with $+q$ placed at the odd lattice site $(1,2)$ and $-q$ placed at the even lattice site $(4,4)$ on a $6\times6$ square lattice. The system is studied at $m/\kappa=12$ and $g_E/\kappa=24$ for different values of the magnetic coupling $g_B/\kappa$ in the $2+1$D spin-$1$ U$(1)$ QLM. (a) The fidelity $\mathcal{F}$ with respect to the initial string state. (b) The total overlap $\mathcal{P}_{\gamma\neq\gamma_{\text{i}}}$ with all minimal string configurations excluding the initial string. (c) The total matter occupation $\langle\hat{n}\rangle$ computed within the minimal patch containing the two static charges.}
\label{fig:charge1_res}
\end{figure}

\begin{figure}
\includegraphics[width=1.0 \linewidth]{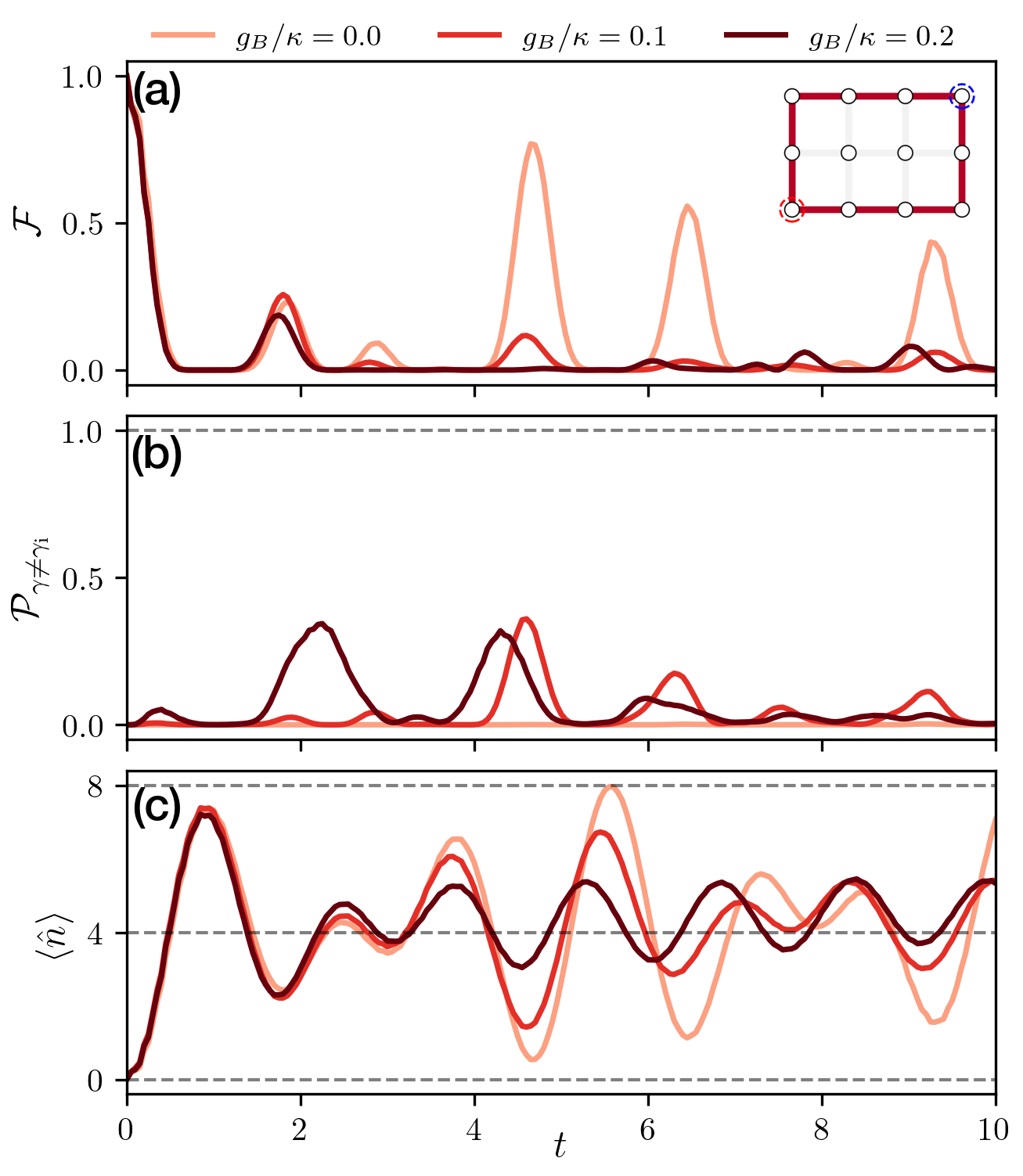}
\caption{Resonant dynamics of a rectangular string (see inset) between static charges of magnitude $q=2$ with $+q$ placed at the odd lattice site $(1,2)$ and $-q$ placed at the even lattice site $(4,4)$ on a $6\times6$ square lattice. The system is studied at $m/\kappa=12$ and $g_E/\kappa=24$ for different values of the magnetic coupling $g_B/\kappa$ in the $2+1$D spin-$1$ U$(1)$ QLM. (a) The fidelity $\mathcal{F}$ with respect to the initial string state. (b) The total overlap $\mathcal{P}_{\gamma\neq\gamma_{\text{i}}}$ with all minimal string configurations excluding the initial string. (c) The total matter occupation $\langle\hat{n}\rangle$ computed within the minimal patch containing the two static charges.}
\label{fig:charge2_res}
\end{figure}

\subsubsection{L-string-breaking dynamics}
We first consider the dynamics of an L-string (see Fig.~\ref{fig:intro_fig}(c)) initialized between static charges $\pm q$ with $q=1$, as shown in Fig.~\ref{fig:charge1_res}. When the plaquette term is switched off ($g_B/\kappa = 0$), we observe near-perfect revivals in both the fidelity $\mathcal{F}$, shown in Fig.~\ref{fig:charge1_res}(a), and the total matter occupation $\langle \hat{n} \rangle$, shown in Fig.~\ref{fig:charge1_res}(c), within the patch. The dynamics is reminiscent of Rabi oscillations in this regime. These revivals are accompanied by a vanishing probability $\mathcal{P}_{\gamma \neq \gamma_{\text{i}}}$, shown in Fig.~\ref{fig:charge1_res}(b), for the string to populate other minimal configurations of the same Manhattan-distance length, which are accessible only in the presence of plaquette terms. Consequently, the string breaks exclusively along its initial configuration, consistent with earlier studies showing that string-breaking dynamics remains effectively $1+1$D in the absence of magnetic plaquette terms \cite{Tian2025RolePlaquetteTerm}. Upon switching on the plaquette term, the fidelity revivals become strongly suppressed with increasing $g_B/\kappa$, accompanied by a finite $\mathcal{P}_{\gamma \neq \gamma_{\text{i}}}$, providing clear evidence of genuine $2+1$D string-breaking dynamics. In parallel, the revivals in the matter density $\langle \hat{n} \rangle$ are likewise suppressed as $g_B/\kappa$ increases, indicating dynamics of a true many-body nature.

\subsubsection{Rectangular string-breaking dynamics}
Interestingly, as discussed before, the $S = 1$ representation of the gauge fields also allows for static charges of magnitude $q = 2$, which are enforced by a vacuum configuration in which the electric fields on all links vanish, and the Gauss's law constraint admits a minimal string configuration that is rectangular in shape, as illustrated in Fig.~\ref{fig:intro_fig}(d). We repeat the quench protocol used for the L-string and study the resulting dynamics, shown in Fig.~\ref{fig:charge2_res}. In the absence of the plaquette term, $g_B/\kappa = 0$,
we observe behavior qualitatively similar to the L-string case where the fidelity $\mathcal{F}$, shown in Fig.~\ref{fig:charge2_res}(a), and the matter density $\langle \hat{n} \rangle$, shown in Fig.~\ref{fig:charge2_res}(c), exhibit revivals in their dynamics, although the revivals are neither perfect nor persistent for the rectangular string. We attribute this to the increased phase space for matter fluctuations surrounding the rectangular string, which provides additional channels for dissipation compared to the L-shaped configuration. Nevertheless, the probability $\mathcal{P}_{\gamma \neq \gamma_{\text{i}}}$ for the string to occupy other minimal configurations, shown in Fig.~\ref{fig:charge2_res}(b), remains essentially zero, confirming that string breaking is effectively $1+1$D, occurring only along the original configuration. Upon turning on the plaquette term, we again observe the expected behavior where string breaking extends beyond the initial configuration. This is manifested by a pronounced suppression of the fidelity revivals, accompanied by a finite $\mathcal{P}_{\gamma \neq \gamma_{\text{i}}}$.

\subsection{Off-resonant string dynamics: Glueballs}
Lastly, we consider the off-resonant regime, where $2m\neq g_E$. Under this condition, string breaking is energetically suppressed during the dynamics, and consequently, no matter is produced from the vacuum. Previous off-resonant studies in QLMs employing an $S=1/2$ representation for the gauge fields have demonstrated that plaquette terms are essential for realizing genuine $2+1$D string dynamics \cite{Tian2025RolePlaquetteTerm}. While this conclusion remains valid for the $S=1$ case, the higher-spin representation enables an additional scenario that is absent in the $S=1/2$ formulation. In particular, the $S=1$ representation allows for the existence of non-minimal string configurations that extend beyond the Manhattan distance, which constituted a bottleneck in the $S = 1/2$ case \cite{Tian2025RolePlaquetteTerm}. This added flexibility makes it possible to prepare highly non-equilibrium initial states, such as the ``snake'' string (see Fig.~\ref{fig:intro_fig}(e)), and to investigate the formation of ``glueballs,'' which are closed Wilson-loop electric flux configurations arising purely from gauge-field dynamics and without accompanying matter creation \cite{Xu2025StringBreakingDynamics,cataldi2025realtimestringdynamics21d}.

\begin{figure}
\includegraphics[width=1.0 \linewidth]{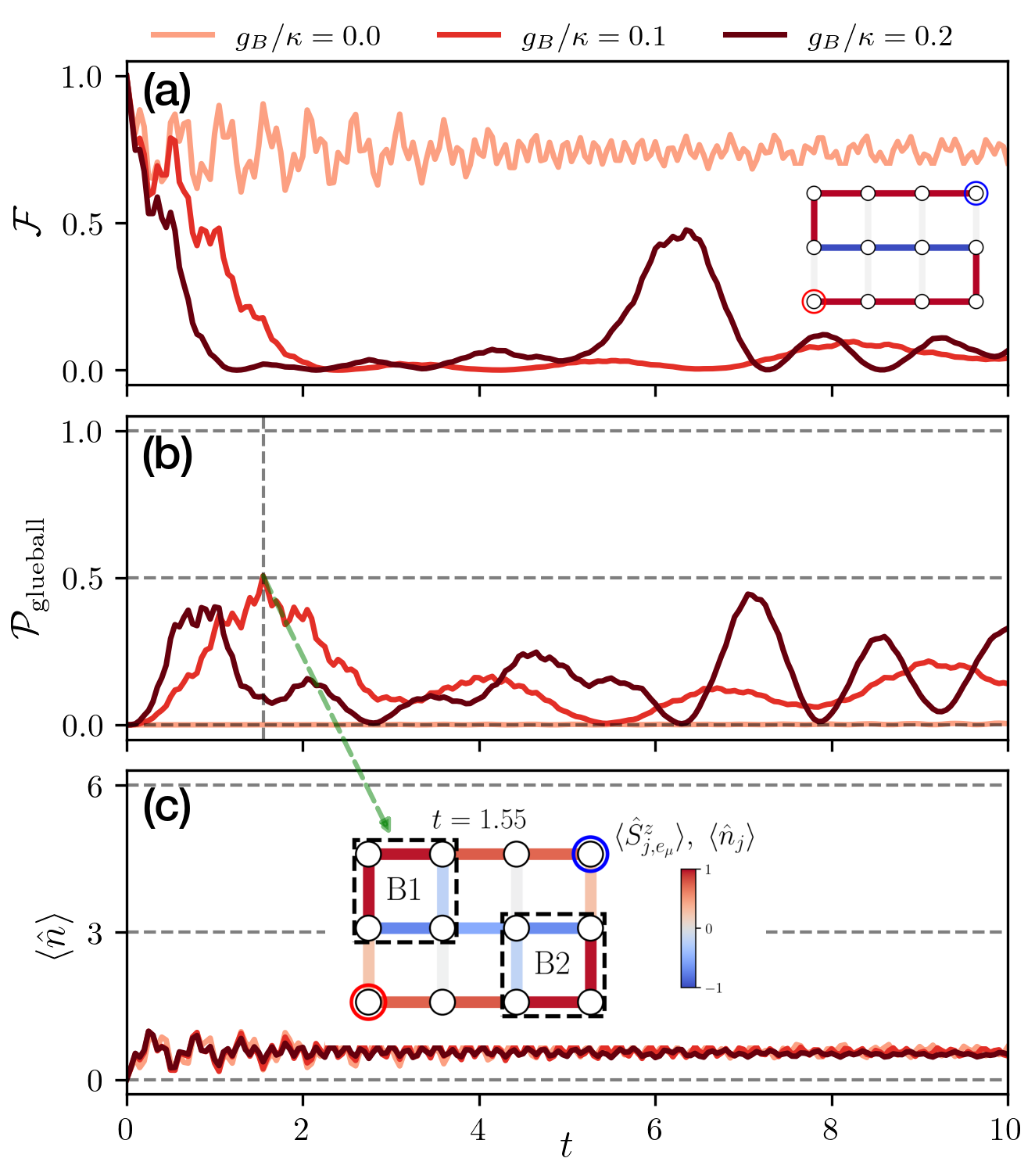}
\caption{Off-resonant string dynamics at $m/\kappa=12$ and $g_E/\kappa=12$ for different values of the magnetic coupling $g_B/\kappa$, starting from a snake-string initial state (see inset) connecting static charges of magnitude $q=1$ with $+q$ placed at the odd lattice site $(1,2)$ and $-q$ placed at the even lattice site $(4,4)$ on a $6\times6$ square lattice in the $2+1$D spin-$1$ U$(1)$ QLM. (a) The fidelity $\mathcal{F}$ with respect to the initial string state. (b) The total overlap $\mathcal{P}_{\textrm{glueball}}$ with all possible bubble configurations (glueballs). (c) The total matter occupation $\langle\hat{n}\rangle$ computed within the minimal patch containing the two static charges. The snapshot at $t=1.55$ displayed in the inset shows the emergence of two glueball configurations labeled B1 and B2.}
\label{fig:charge1_glueball}
\end{figure}

To probe this, we study off-resonant dynamics starting from a snake string as shown in Fig.~\ref{fig:charge1_glueball}. We first consider the case in which the plaquette term is switched off ($g_B = 0$). In this regime, we observe a nonzero fidelity with small fluctuations, shown in Fig.~\ref{fig:charge1_glueball}(a), accompanied by negligible matter fluctuations, as evidenced by the small value of the matter density $\langle \hat{n} \rangle$, shown in Fig.~\ref{fig:charge1_glueball}(c). This indicates that the string neither breaks nor propagates beyond its original configuration. In contrast, turning on the plaquette term ($g_B \neq 0$) leads to a qualitatively different behavior where the fidelity drops significantly, and a substantial nonzero probability $\mathcal{P}_{\mathrm{glueball}}$, shown in Fig.~\ref{fig:charge1_glueball}(b), emerges for the system to occupy glueball configurations (see the inset of Fig.~\ref{fig:charge1_glueball}(c)), while the matter density remains close to zero. This therefore represents an intriguing regime in which string breaking occurs without accompanying matter creation.

\section{Quantum simulation with digital qudit processors}

For higher-spin ($S > 1/2$) QLMs, the intrinsic multilevel structure of qudits \cite{Ringbauer_2022} offers a natural platform for implementation, where different spin states can be encoded in distinct energy levels. Building on this idea, we propose an explicit realization for our string-breaking experiments in the spin-$1$ case by leveraging the qudit structure provided by a trapped-ion qudit architecture.

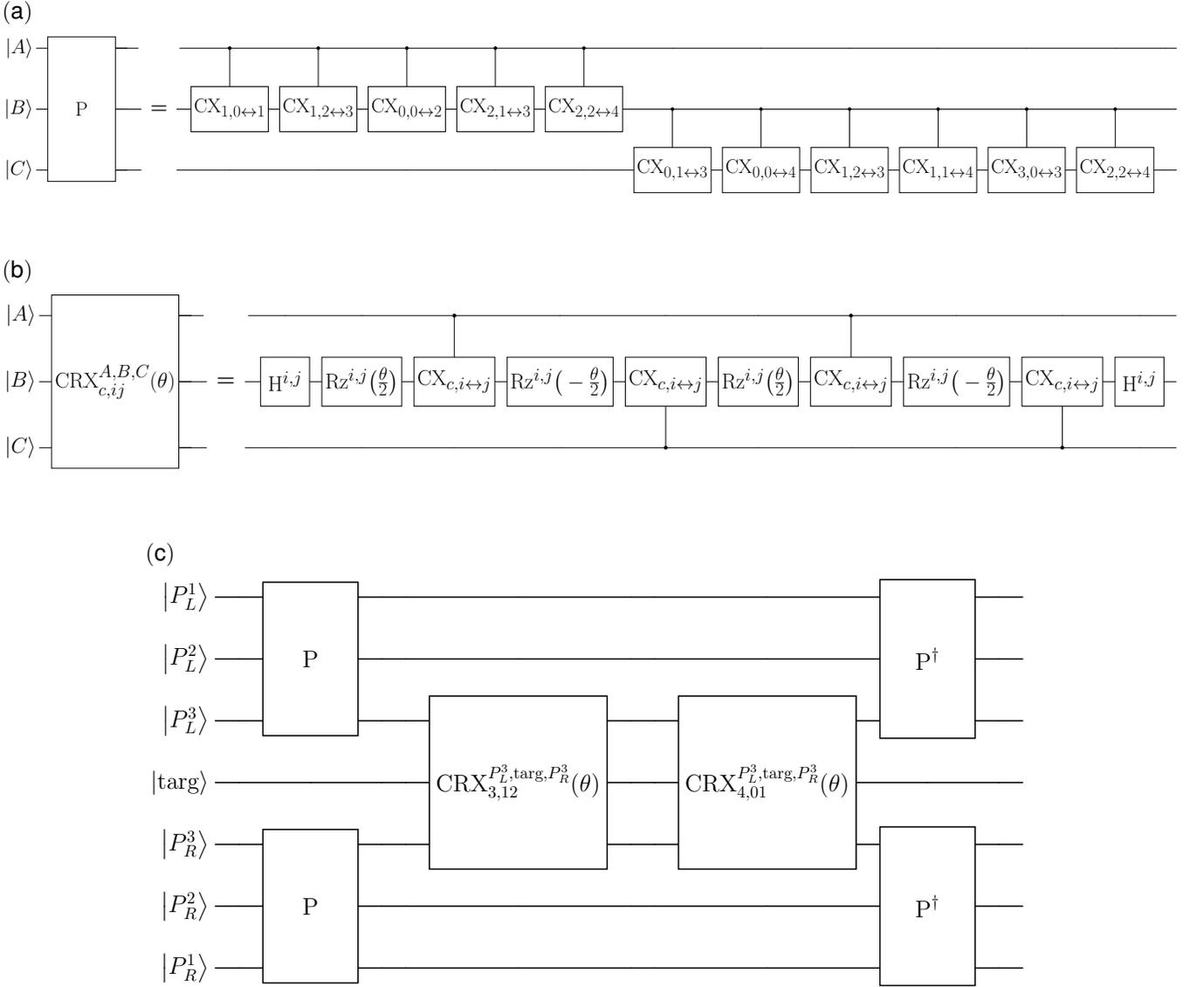
\begin{figure*}[t]
\captionsetup[subfloat]{position=top,justification=raggedright,singlelinecheck=false, font=normalsize}
  \centering
  \subfloat[]{%
    \resizebox{\linewidth}{!}{%
\begin{quantikz}[row sep={2.7cm,between origins}, font=\Huge]
    \lstick{$\ket{A}$} & \gate[3][3cm][1cm]{\text{P}} & \qw & \midstick[3,brackets=none]{\textbf{\Huge =}} & \ctrl{1} & \ctrl{1} & \ctrl{1} & \ctrl{1} & \ctrl{1} & &  & & & & & &\\
    \lstick{$\ket{B}$} & & \qw & & \gate[1][2cm][2cm]{\text{CX}_{1, 0\leftrightarrow 1}} & \gate[1][2cm][2cm]{\text{CX}_{1, 2\leftrightarrow 3}} & \gate[1][2cm][2cm]{\text{CX}_{0, 0\leftrightarrow 2}} & \gate[1][2cm][2cm]{\text{CX}_{2, 1\leftrightarrow 3}} & \gate[1][2cm][2cm]{\text{CX}_{2, 2\leftrightarrow 4}} & \ctrl{1} & \ctrl{1} &\ctrl{1} & \ctrl{1} & \ctrl{1} & \ctrl{1} & &\\
    \lstick{$\ket{C}$} & & \qw & & & & & & & \gate[1][2cm][2cm]{\text{CX}_{0, 1\leftrightarrow 3}} & \gate[1][2cm][2cm]{\text{CX}_{0, 0\leftrightarrow 4}} & \gate[1][2cm][2cm]{\text{CX}_{1, 2\leftrightarrow 3}} &\gate[1][2cm][2cm]{\text{CX}_{1, 1\leftrightarrow 4}} & \gate[1][2cm][2cm]{\text{CX}_{3, 0\leftrightarrow 3}} & \gate[1][2cm][2cm]{\text{CX}_{2, 2\leftrightarrow 4}} & &\\
\end{quantikz}
    }%
  }\hfill
  \subfloat[]{%
    \resizebox{\linewidth}{!}{%
    \begin{quantikz}[row sep={2.7cm,between origins}, font=\Huge]
    \lstick{$\ket{A}$} & \gate[3][4cm][1cm]{\text{CRX$^{A,B,C}_{c, ij}$}(\theta)} & \qw & \midstick[3,brackets=none]{\textbf{\Huge =}} & & & \ctrl{1} & & & &  \ctrl{1} &  & & &\\
    \lstick{$\ket{B}$} & & \qw & & \gate[1][2cm][2cm]{\text{H}^{i,j}} &  \gate[1][2cm][2cm]{\text{Rz}^{i,j}\big(\frac{\theta}{2}\big)} & \gate[1][2cm][2cm]{\text{CX}_{c, i \leftrightarrow j}}  & \gate[1][2cm][2cm]{\text{Rz}^{i,j}\big(-\frac{\theta}{2}\big)} & \gate[1][2cm][2cm]{\text{CX}_{c, i \leftrightarrow j}} & \gate[1][2cm][2cm]{\text{Rz}^{i,j}\big(\frac{\theta}{2}\big)} & \gate[1][2cm][2cm]{\text{CX}_{c, i \leftrightarrow j}} & \gate[1][2cm][2cm]{\text{Rz}^{i,j}\big(-\frac{\theta}{2}\big)} & \gate[1][2cm][2cm]{\text{CX}_{c, i \leftrightarrow j}} & \gate[1][2cm][2cm]{\text{H}^{i,j}} & \\
    \lstick{$\ket{C}$} & & \qw & & & & & &  \ctrl{-1} & & & & \ctrl{-1} & & \\
\end{quantikz}
    }%
  }
  \hfill
  \subfloat[]{%
    \resizebox{0.75\linewidth}{!}{%
        \begin{quantikz}[row sep={1.3cm,between origins}, font=\Large]
    \lstick{$\ket{P_L^1}$} & & \gate[3][2cm]{\text{P}} & & & & & & & \gate[3][2cm]{\text{P}^{\dag}} & &\\
    \lstick{$\ket{P_L^2}$} & & & & & & & & & & & \\
    \lstick{$\ket{P_L^3}$} & & & & & \gate[3][2cm]{\text{CRX$^{P_L^3, \text{targ}, P_R^3}_{3,12}$}(\theta)} & & & \gate[3][2cm]{\text{CRX$^{P_L^3, \text{targ}, P_R^3}_{4, 01}$}(\theta)} & & & \\
    \lstick{$\ket{\text{targ}}$} & & & & & & & & & & & \\
    \lstick{$\ket{P_R^3}$} & & \gate[3][2cm]{\text{P}} & & & & & & & \gate[3][2cm]{\text{P}^{\dag}} & & \\
    \lstick{$\ket{P_R^2}$} & & & & & & & & & & & \\
    \lstick{$\ket{P_R^1}$} & & & & & & & & & & & \\
    \end{quantikz}

}
}
    \caption{(a) The projector verification subcircuit $\mathrm{P}$ maps 3-qudit states according to Tables~\ref{tab:s_1_p-1} and~\ref{tab:s_1_p0}. (b) The subcircuit $\mathrm{CRX}^{A,B,C}_{ c, ij}$($\theta$) performs a controlled-X rotation by an angle $\theta$ within the $\{\ket{i}, \ket{j}\}$ subspace, conditioned on both the control qudits being in $\ket{c}$. The control qudits are labeled by index $A$, $C$ and the target qudit by $B$. (c) Full quantum circuit for implementing evolution $\hat{U}_{C_{\mathbf{j, e_{\mu}}}}^{\rm MIO} (\theta)$:  The complete quantum circuit includes 52 two-qudit gates and 12 single-qudit gates, with a two-qudit gate depth of 30.
}
\label{fig:U_c_spin_1}
\end{figure*}

\begin{figure*}[t]
    \resizebox{\linewidth}{!}{%
      \begin{quantikz}[row sep={2cm,between origins}, font=\Large]
    \lstick{$\ket{A}$} & \gate[4][2.5cm][1cm]{\mathrm{exp}(-\imath\theta X_A X_B X_C X_D)} & \qw & \midstick[4,brackets=none]{\textbf{\large =}} & & \gate[4][2cm][1cm]{\mathrm{MS}^{01}_{A, D}(\alpha, \beta)} & & & & & &  \gate[4][2cm][1cm]{\mathrm{MS}^{01}_{A, D}(\alpha, \beta)} & &\\
    \lstick{$\ket{B}$} & & \qw & & & &  \gate[3][2cm][1cm]{\mathrm{MS}^{01}_{B, D}(\alpha, \beta)} & & & &  \gate[3][2cm][1cm]{\mathrm{MS}^{01}_{B, D}(\alpha, \beta)} & & &  \\
    \lstick{$\ket{C}$} & & \qw & & & & &  \gate[2][2cm][1cm]{\mathrm{MS}^{01}_{C, D}(\alpha, \beta)} & &  \gate[2][2cm][1cm]{\mathrm{MS}^{01}_{C, D}(\alpha, \beta)} & & & & \\
    \lstick{$\ket{D}$} & & \qw & & \gate[1][2cm][2cm]{\text{Rz}^{01} (\frac{\pi}{2})} & & & & \gate[1][2cm][2cm]{\text{Rz}^{01} (\pi - 2\theta)} &  & & & \gate[1][2cm][2cm]{\text{Rz}^{01} (\frac{\pi}{2})} & \\
\end{quantikz}
}
\caption{The circuit implementing the term $e^{-\imath \theta \, \hat{\sigma}^{x;01}_{\mathbf{j},1} \hat{\sigma}^{x;01}_{\mathbf{j},2} \hat{\sigma}^{x;01}_{\mathbf{j},3} \hat{\sigma}^{x;01}_{\mathbf{j},4}}
$ through decomposition into fixed-angle MS gates and local $\operatorname{Rz}^{ab}(\theta)$ gates.}\label{fig:XXXX_1}
\end{figure*}
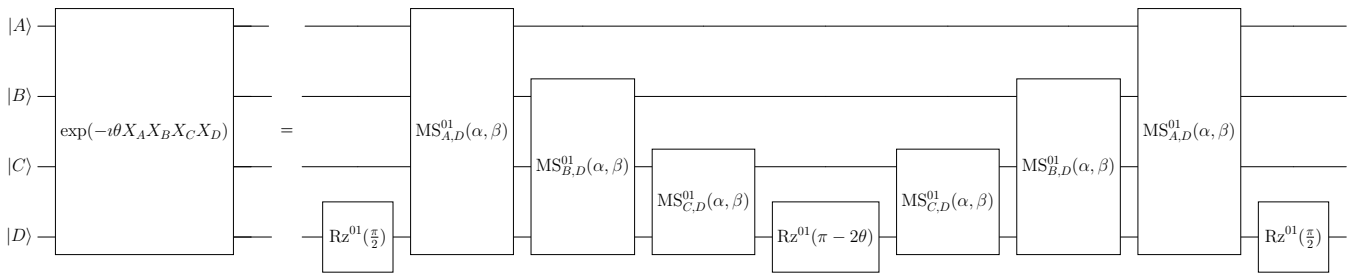

By incorporating a particle-hole transformation \cite{Hauke2013}, Gauss's law can be recast in a form that explicitly relates the matter occupation at a lattice site to the sum of electric fields on the surrounding links. The matter fields can then be integrated out systematically in the $g_{\mathbf{j}} = 0$ sector, yielding an equivalent Hamiltonian expressed solely in terms of gauge degrees of freedom \cite{joshi2025efficientquditcircuitquench}. Within this framework, a static charge $q$ at any site $\mathbf{j}$ can be equivalently defined by manipulating the links around that site. 

For the $S=1$ representation, the resulting matter-integrated-out (MIO) Hamiltonian takes the form,

\begin{equation}\label{h_qudit_qlm}
\begin{aligned}
\hat{H}_{\mathrm{MIO}} ={}& -\kappa 
 \Big (\sum_{\mathbf{j}, \mathbf{\mu}} \hat{P}^{-1}_{\mathbf{j}} \, \hat{\sigma}^{x;0, 1}_{\mathbf{j}, \mathbf{e_{\mu}}} \, \hat{P}^{-1}_{\mathbf{j + e_{\mu}}} + \sum_{\mathbf{j}, \mathbf{\mu}} \hat{P}^{0}_{\mathbf{j}} \, \hat{\sigma}^{x;0, -1}_{\mathbf{j}, \mathbf{e_{\mu}}} \, \hat{P}^{0}_{\mathbf{j + e_{\mu}}} \Big)\\
& - 2m \sum_{\mathbf{j}, \mathbf{\mu}} \hat{S}^z_{\mathbf{j}, \mathbf{e_{\mu}}}  + g_E \sum_{\mathbf{j}, \mathbf{\mu}} (\hat{S}^z_{\mathbf{j}, \mathbf{e_{\mu}}})^2 \\
& + g_B \sum_{\mathbf{j}}\left(\hat{U}_{\square}+\hat{U}_{\square}^{\dagger}\right),
\end{aligned}
\end{equation}
where the projectors $\hat{P}^{m}_{\mathbf{j}}$ and $\hat{P}^{m}_{\mathbf{j+\mathbf{e_\mu}}}$ are defined over all the links surrounding the matter sites $\mathbf{j}$ and $\mathbf{j+\mathbf{e_\mu}}$, respectively, excluding $(\mathbf{j}, \mathbf{e_\mu})$. The original prefactor $\sqrt{S(S+1)-m(m+1)}$ has been absorbed into the coupling strength $\kappa$.
The operator $\hat{\sigma}^{x;-m, -m-1}_{\mathbf{j}, \mathbf{e_{\mu}}}$ is the flip operator on the link ($\mathbf{j}, \mathbf{e_{\mu}}$) within the two-dimensional subspace spanned by $\{\ket{-m}, \ket{-m-1}\}$, conditioned on the satisfaction of both projectors $\hat{P}^{m}_{\mathbf{j}}$ and $\hat{P}^{m}_{\mathbf{j + e_{\mu}}}$ on its neighboring links. Finally, the plaquette operator is given by $\hat{U}_{\square} = \hat{U}_{\mathbf{j},\mathbf{e}_x}\hat{U}^\dagger_{\mathbf{j}+\mathbf{e}_x,\mathbf{e}_y}
\hat{U}_{\mathbf{j}+\mathbf{e}_y,\mathbf{e}_x}
\hat{U}^\dagger_{\mathbf{j},\mathbf{e}_y}$.

To simulate time evolution under $\hat{H}_{\mathrm{MIO}}$, we make use of the first-order Suzuki-Trotter decomposition,
\begin{equation}\label{u_trotter}
\begin{aligned}
\hat{U}_{\rm ST}(\theta)
\;=&\;
\Biggl(
\prod_{\mathbf{j}, \mathbf{\mu}}
\underbrace{e^{\imath 2m\hat{S}_{\mathbf{j, e_{\mu}}}^z d\theta}}_{\displaystyle \hat{U}_{M_{\mathbf{j, e_{\mu}}}}^{\rm MIO} (\theta)}\,
\underbrace{e^{-\imath g_E(\hat{S}_{\mathbf{j, e_{\mu}}}^z)^2 d\theta}}_{\displaystyle \hat{U}_{E_{\mathbf{j, e_{\mu}}}}^{\rm MIO} (\theta)}\\
&\underbrace{e^{\imath \kappa \sum_{m = 0, -1} \hat{P}_{\mathbf{j}}\, \hat{\sigma}^{x;-m, -m-1}_{\mathbf{j, e_{\mu}}} \hat{P}_{\mathbf{j+ e_{\mu}}} d\theta}}_{\displaystyle \hat{U}_{C_{\mathbf{j, e_{\mu}}}}^{\rm MIO}(\theta)}\,
\underbrace{e^{- \imath g_B \hat{H}_{\square} d\theta}}_{\displaystyle \hat{U}_{\square}^{\rm MIO}(\theta)}
\Biggr)^{N},
\end{aligned}
\end{equation}
where $d\theta = \theta/N$ denotes the size of a single Suzuki-Trotter step. We encode each spin-$1$ link into qudits with a maximum local Hilbert space dimension of $5$, where two additional levels are used towards constructing efficient circuits. Our notation assigns the three-link states as $ \ket{0} $ for $ S^z = -1 $, $ \ket{1} $ for $ S^z = 0 $, and $ \ket{2} $ for $ S^z = 1 $. The remaining qudit states are labeled as $\ket{3}$ and $\ket{4}$. The first and second unitary operators, $\hat{U}_{M_{\mathbf{j, e_{\mu}}}}^{\rm MIO} (\theta)$ and $\hat{U}_{E_{\mathbf{j, e_{\mu}}}}^{\rm MIO} (\theta)$ in $\hat{U}_{\mathrm{ST}} (\theta)$ \eqref{u_trotter} are single-qudit operations, and can be easily implemented using virtual phase shift operations
\begin{equation}
\operatorname{VRz}^{a}(\phi) = e^{-\imath\phi \ketbra{a}} \label{virz}
\end{equation}
within the $\{\ket{0}, \ket{1}\, \ket{2}\}$ subspace of each qudit. These operations can be realized through classical frame updates directly, rather than physical pulses, and are therefore noiseless.

The circuits for $\hat{U}_{C_{\mathbf{j, e_{\mu}}}}^{\rm MIO} (\theta)$ are shown in Fig.~\ref{fig:U_c_spin_1}, constructed by following the scheme based on the mapping table approach proposed in \cite{joshi2025efficientquditcircuitquench}. For completeness, we provide additional background details on this construction in Appendix~\ref{Sec:Circs_coupling}, where we also present the explicit mapping tables used in our implementation. This approach efficiently implements $\hat{U}_{C_{\mathbf{j, e_{\mu}}}}^{\rm MIO} (\theta)$ using a total of 52 entangling gates for the spin-$1$ QLM; cf.~Fig.~\ref{fig:U_c_spin_1}.
The corresponding circuits are composed entirely of single-qudit operations, and controlled exchange (CX) gates \cite{Ringbauer_2022}, 
\begin{equation}\label{CX}
\begin{aligned}
\operatorname{CX}_{c, l_1 \leftrightarrow l_2}:\left\{
\begin{array}{l}
\ket{c, l_1} \leftrightarrow \ket{c, l_2} \\
\ket{j, k} \rightarrow \ket{j, k} \quad \text{for } j \neq c, \, k \neq l_1, l_2,
\end{array}
\right. 
\end{aligned}
\end{equation} 
which applies $\sigma^{x; l_1, l_2}$ on the target qudit, conditioned on the control qudit being in state $\ket{c}$.

We also provide details for the construction of the circuits for the plaquette term in Appendix \ref{Sec:Circs_plaquette}. The fundamental building block for implementing $\hat{U}_{\square}^{\rm MIO} (\theta)$ is the circuit realizing
$e^{-\imath \theta \, \hat{\sigma}^{x;01}_{\mathbf{j},1} \, \hat{\sigma}^{x;01}_{\mathbf{j},2} \, \hat{\sigma}^{x;01}_{\mathbf{j},3} \, \hat{\sigma}^{x;01}_{\mathbf{j},4}}$,
where $(\mathbf{j},l)$ with $l=1,2,3,4$ labels the four links involved in the plaquette, namely  $(\mathbf{j},\mathbf{e}_x)$, $(\mathbf{j}+\mathbf{e}_x,\mathbf{e}_y)$, $(\mathbf{j}+\mathbf{e}_y,\mathbf{e}_x)$, and $(\mathbf{j},\mathbf{e}_y)$, respectively. The corresponding circuit, shown in Fig.~\ref{fig:XXXX_1}, requires six Mølmer–Sørensen (MS) gates \eqref{eq:MS} and a few noiseless single-qudit operations.

It should be noted that the above construction assumes fixed-angle MS gates and thus, the parameter $\theta$ is controlled through a single local noiseless operation. Alternatively,  $\theta$ can be controlled directly with the MS gates themselves, which allows for a reduced gate count, as illustrated by the decomposition in Eq.~\eqref{plaq_decomp_3}. 

Since all 128 exponential terms involved in the Trotter decomposition (\ref{plaq_decomp_1}, \ref{plaq_decomp_2}) of $\hat{U}_{\square}^{\rm MIO} (\theta)$ are composed exclusively of $\hat{\sigma}^{x;ab}$ and $\hat{\sigma}^{y;ab}$, the circuits for all remaining terms can be directly generated from this fundamental block by supplementing it with local basis transformations on appropriate qudits. As an illustrative example, one may write 
\begin{equation}
\begin{aligned}
e^{-\imath \theta \, \hat{\sigma}^{x;01}_{\mathbf{j},1} \hat{\sigma}^{x;01}_{\mathbf{j},2} \hat{\sigma}^{x;01}_{\mathbf{j},3} \hat{\sigma}^{x;12}_{\mathbf{j},4}}
= & \left[ \mathbb{1} \otimes \mathbb{1} \otimes \mathbb{1} \otimes \hat{\sigma}^{x;02} \right] \\
& e^{-\imath \theta \, \hat{\sigma}^{x;01}_{\mathbf{j},1} \hat{\sigma}^{x;01}_{\mathbf{j},2} \hat{\sigma}^{x;01}_{\mathbf{j},3} \hat{\sigma}^{x;01}_{\mathbf{j},4}} \\
& \left[ \mathbb{1} \otimes \mathbb{1} \otimes \mathbb{1} \otimes \hat{\sigma}^{x;02} \right].
\end{aligned}
\end{equation}
A naive gate-counting estimate would suggest a requirement of 768 entangling gates using the fixed-angle MS gate construction (Fig.~\ref{fig:XXXX_1}), or 640 entangling gates using the tunable MS gate decomposition of Eq.~\eqref{plaq_decomp_3}. However, by employing an optimal ordering of the exponential term in the Trotter decomposition, a subset of these entangling gates cancels between consecutive terms, resulting in a reduced overall gate count and circuit depth.

\section{Conclusion and outlook}

In this work, we investigated string breaking in $2+1$D lattice QED using the quantum link model with spin-$1$ representation of the gauge fields. By employing extensive tensor network methods, we studied string breaking both in and out of equilibrium in the confined regime. In the equilibrium setting, we analyzed string breaking in the ground state induced by static charges for different values of their magnitude. At large values of the latter, a two-stage string-breaking mechanism emerged that is absent in the spin-$1/2$ QLM. Out of equilibrium, we investigated resonant string breaking following quenches from various string configurations, including the rectangular string, which is likewise inaccessible in the spin-$1/2$ formulation for reasonable choice of static charges. We showed that the plaquette term plays a crucial role in enabling genuine $2+1$D string dynamics, in agreement with recent studies \cite{Tian2025RolePlaquetteTerm}. Furthermore, the spin-$1$ representation allowed for the preparation of highly nonequilibrium strings exceeding the minimal length between the two static charges, and we demonstrated that their off-resonant dynamics leads to glueball formation that cannot arise in the spin-$1/2$ representation. Finally, we proposed efficient qudit circuit implementations for ion-trap quantum simulators, demonstrating the practical realizability of the proposed string dynamics.

Building on our findings here, we see several avenues worth exploring. For example, it is important to note that the relevance of U$(1)$ QLMs extends far beyond the scope of string breaking. They have recently become a major platform for nonergodic phenomena, in particular quantum many-body scarring \cite{Bernien2017ProbingManyBodyDynamics,Turner2018WeakErgodicityBreaking,Biswas2022ScarsFromProtectedZeroModes,Serbyn2021QuantumManyBodyScars,Moudgalya2022QuantumManyBodyScarsHilbertSpaceFragmentation,Chandran2023QuantumManyBodyScars,Bluvstein2022QuantumProcessor,Desaules2023WeakErgodicityBreaking,Desaules2023ProminentQuantumManyBodyScars,Zhang2023ManyBodyHilbertSpaceScarring,Dong2023DisorderTunableEntanglement,Osborne2024QuantumManyBodyScarring,Budde2024QuantumManyBodyScars}. The latter also has been shown to have a direct connection to string breaking \cite{Surace2020LatticeGaugeTheories} and plasma dynamics \cite{mark2025observationballisticplasmamemory} in one spatial dimension and for $S=\frac{1}{2}$. As such, it would be interesting to study the effect of higher-level representations of the gauge field and two spatial dimensions on plasma dynamics and quantum many-body scarring in the $2+1$D U$(1)$ QLM. Another route to explore is whether going to even larger values of $S$ will lead to any qualitatively different behavior in string breaking. It may be that larger $S$ can lead to a multi-stage string-breaking mechanism in equilibrium, so it would be worth exploring this to connect it to the string-breaking behavior in the Kogut--Susskind limit. 

On the quantum simulation side, it would be interesting to explore analog cold-atom quantum simulators, which have been demonstrated theoretically in $2+1$D for $S=\frac{1}{2}$ \cite{Osborne2025Scale$2+1$DGaugeTheory} and in $1+1$D for $S=1$ \cite{osborne2023spinsmathrmu1quantumlink}. In principle, both these frameworks can be combined to realize a cold-atom quantum simulator of a $2+1$D spin-$1$ U$(1)$ QLM, and where ring-exchange interactions can be employed to engineer a plaquette term \cite{Dai2017FourBodyRingExchange}.

\bigskip

\footnotesize
\begin{acknowledgments}
    The authors acknowledge fruitful discussions with Debasish Banerjee, Giovanni Cataldi, Jesse J.~Osborne, and Torsten V.~Zache. The authors acknowledge funding by the Max Planck Society, the Deutsche Forschungsgemeinschaft (DFG, German Research Foundation) under Germany’s Excellence Strategy – EXC-2111 – 390814868, and the European Research Council (ERC) under the European Union’s Horizon Europe research and innovation program (Grant Agreement No.~101165667)—ERC Starting Grant QuSiGauge. Views and opinions expressed are however those of the author(s) only and do not necessarily reflect those of the European Union or the European Research Council Executive Agency. Neither the European Union nor the granting authority can be held responsible for them. This work is part of the Quantum Computing for High-Energy Physics (QC4HEP) working group.
\end{acknowledgments}
\normalsize

\appendix

\section{Convergence tests}\label{app:convergence}

\begin{figure*}
\centering
\includegraphics[width=1.0 \linewidth]{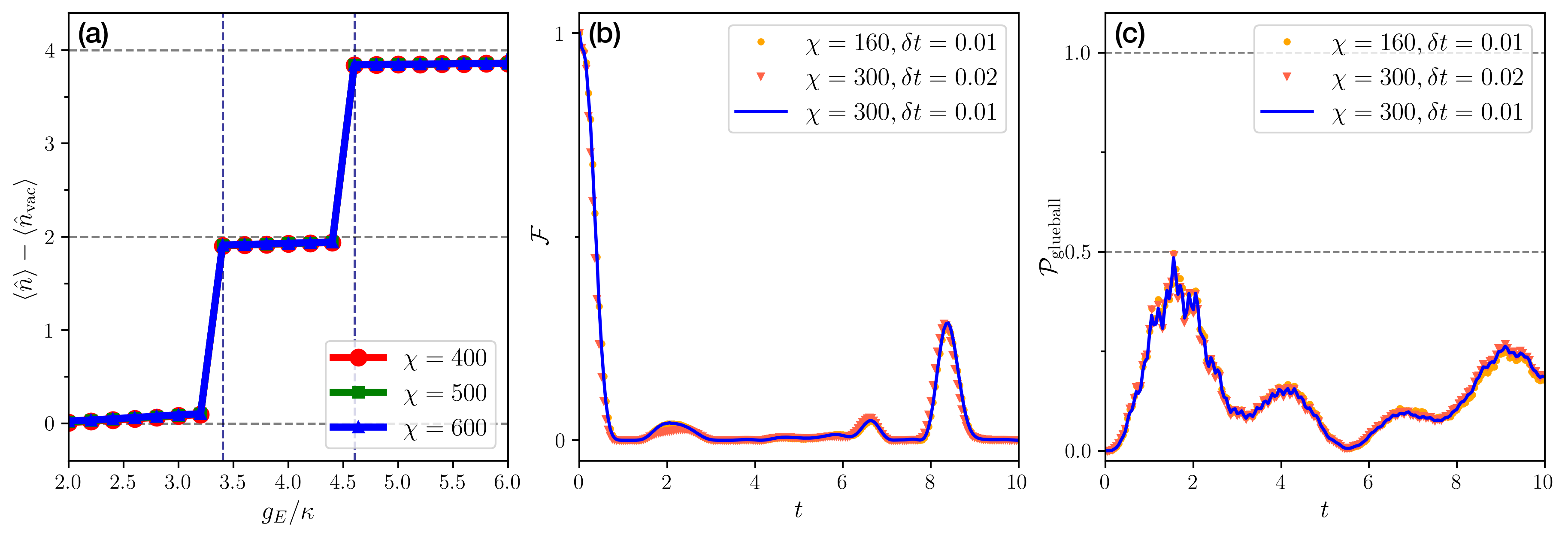}
\caption{Convergence analysis of different observables with respect to the MPS bond dimension $\chi$ and the TDVP time step $\delta t$ across representative parameter regimes in both equilibrium and nonequilibrium dynamics of the $2+1$D spin-$1$ U$(1)$ QLM. We focus on the most demanding cases, requiring the largest $\chi$ and smallest $\delta t$ to achieve convergence. (a) Vacuum-subtracted matter density $\langle \hat{n} \rangle - \langle \hat{n}_{\mathrm{vac}} \rangle$ as a function of the electric field strength $g_E$ for the case with two static charges of magnitude $q=2$ and with $g_B/\kappa=0.1$. (b) Fidelity $\mathcal{F}$ for rectangular string-breaking dynamics at resonance ($2m=g_E$) with magnetic field strength $g_B/\kappa=0.2$. (c) Glueball probability $\mathcal{P}_{\mathrm{glueball}}$ at $g_B/\kappa=0.1$ for snake-string dynamics off resonance at $m/\kappa=12$ and $g_E/\kappa=12$.
}
\label{fig:convergence}
\end{figure*}

To assess the accuracy of our simulations, which rely on matrix product state (MPS) approximations for the ground-state calculations using DMRG and for the real-time evolution using TDVP, we perform a detailed convergence analysis with respect to the MPS bond dimension $\chi$ and the time step $\delta t$. Fig.~\ref{fig:convergence} presents the most demanding cases encountered in the main text, including equilibrium string breaking between opposite static charges of magnitude $q=2$, the fidelity of resonant rectangular string-breaking dynamics at finite magnetic coupling, and off-resonant glueball dynamics initialized from a snake string configuration. In all cases, we observe good convergence, thereby validating the accuracy and reliability of our numerical simulations.

\section{Alternative scenario for the placement of static charges} \label{alternate}

We now consider an alternate scenario in which a negative static charge is enforced on an odd lattice site and a positive static charge on an even site. We repeat the equilibrium string-breaking analysis presented in the main text for this complementary static charge configuration and compare it with the case studied in the main text.

\begin{figure}
\centering
\includegraphics[width=1.0 \linewidth]{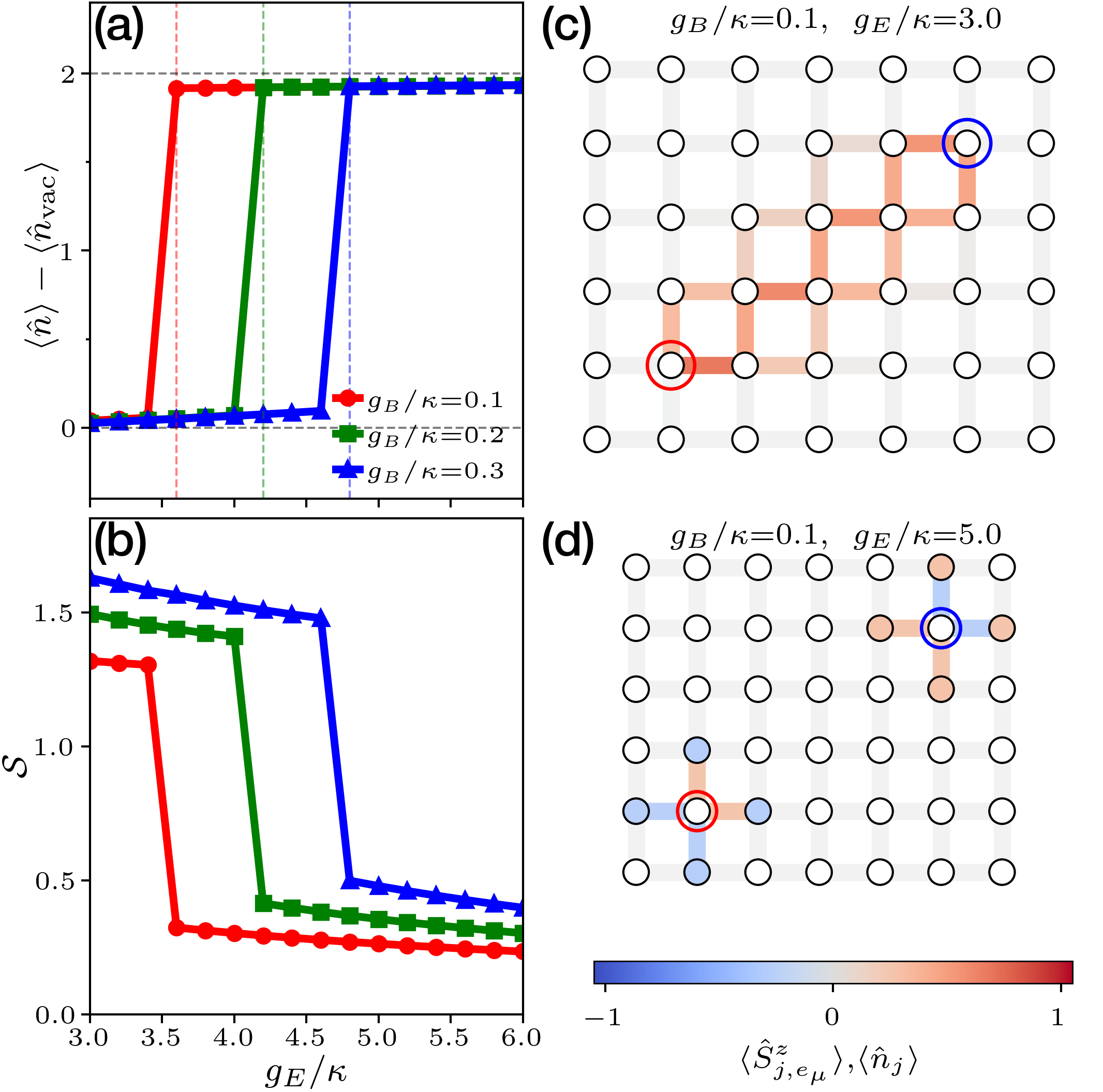}
\caption{Equilibrium string breaking between static charges of magnitude $q=1$ with $+q$ placed at the even lattice site $(5,1)$ and $-q$ placed at the odd lattice site $(9,4)$ on a rectangular lattice of dimension $L_x=16$ and $L_y=6$ for the $2+1$D spin-$1$ U$(1)$ QLM as the electric coupling $g_E$ is increased at fixed $g_B$. String breaking is signaled by (a) a sharp increase in the vacuum-subtracted matter density $\langle \hat{n} \rangle - \langle \hat{n}_{\mathrm{vac}} \rangle$, corresponding to the creation of a positron-electron pair, and (b) a rapid drop in the entanglement entropy $\mathcal{S}$. A finite magnetic coupling $g_B$ stabilizes the confining string, shifting the breaking point to larger values of $g_E/\kappa$. (c) Representative string state (showing only the central patch containing static charges) prior to breaking, computed at $g_E/\kappa=3.0$ and $g_B/\kappa = 0.1$, where the two static charges are connected by a superposition of electric flux string configurations. (d) Broken-string state at $g_E/\kappa = 5.0$ and $g_B/\kappa = 0.1$, in which the string disappears and an additional electron is distributed over the four odd lattice sites surrounding the lattice site $(5,1)$, while the additional positron is distributed over the four even lattice sites surrounding the lattice site $(9,4)$. For all the cases, we use $m = 6$.}
\label{fig:SMcharge1}
\end{figure}

For the $q=1$ case, a qualitatively similar string-breaking transition is observed, as evidenced by the characteristic jumps in the total vacuum-subtracted matter density, $\langle \hat{n} \rangle - \langle \hat{n}_{\mathrm{vac}} \rangle$, and in the entanglement entropy, \(\mathcal{S}\), shown in Fig.~\ref{fig:SMcharge1} (a,b). We further observe that increasing the magnetic strength $g_B$ stabilizes the string, shifting the breaking transition to larger $g_E$, consistent with the behavior discussed in the main text. Importantly, however, unlike the configuration analyzed in Fig.~\ref{fig:combined_charge1} of the main text, the staggered fermion formulation in this case does not permit the creation of dynamical matter on the same lattice sites as the static charges. The reason is as follows: with a $+q$ static charge enforced on an even lattice site, screening can only occur by populating electrons on the four odd sites surrounding the static charge. Similarly, to screen the $-q$ static charge on an odd lattice site, the positron can only be placed on the four surrounding even sites. This effect is clearly visible in the snapshot shown in Fig.~\ref{fig:SMcharge1}(d).

\begin{figure}
\centering
\includegraphics[width=1.0 \linewidth]{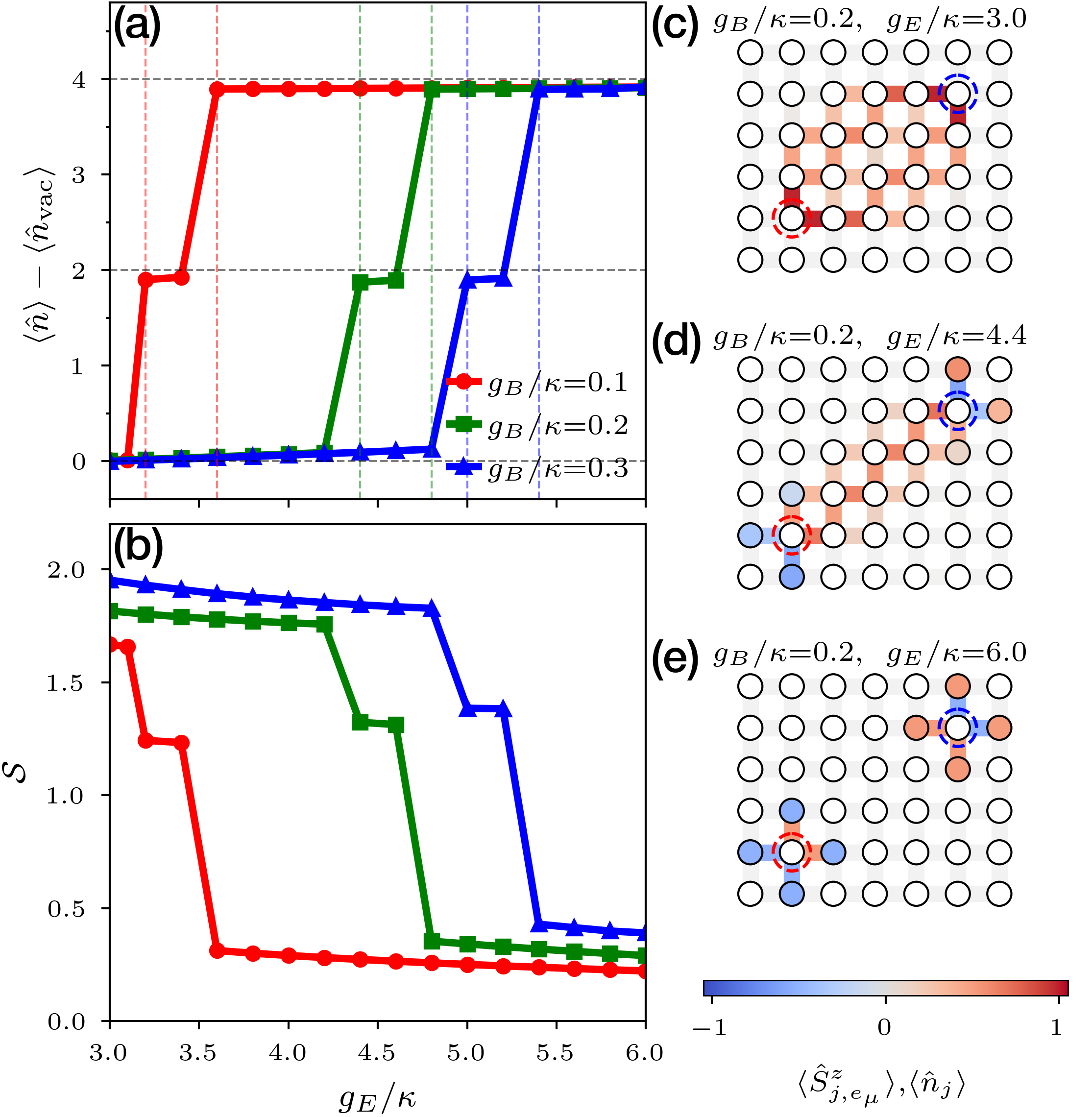}
\caption{Equilibrium string breaking between static charges of magnitude $q=2$ with $+q$ placed at the even lattice site $(5,1)$ and $-q$ placed at the odd lattice site $(9,4)$ on a rectangular lattice of dimension $L_x=16$ and $L_y=6$ for the $2+1$D U$(1)$ QLM as the electric coupling $g_E$ is increased at fixed $g_B$. The two-stage string breaking process is reflected in (a) two successive jumps in the vacuum-subtracted matter density $\langle \hat{n} \rangle - \langle \hat{n}_{\mathrm{vac}} \rangle$, each corresponding to the creation of a positron-electron pair, and (b) two corresponding jumps in the entanglement entropy $\mathcal{S}$. A finite magnetic coupling $g_B$ stabilizes the confining strings, shifting both breaking points to larger values of $g_E/\kappa$. (c) Representative string state (showing only the central patch containing static charges) prior to breaking, computed at $g_E/\kappa = 3.0$ and $g_B/\kappa  = 0.2$, where the static charges are connected by a superposition of electric flux string configurations. (d) Intermediate state after the first string breaking event, in which an electron is distributed over the four odd lattice sites surrounding the lattice site $(5,1)$, while the positron is distributed over the four even lattice sites surrounding the lattice site $(9,4)$, while the static charges remain connected by a residual string. (e) Final broken-string state after the second breaking event, where the remaining string completely disappears and additional electron and a positron is created at the sites surrounding the lattice sites $(5,1)$ and $(9,4)$ respectively. All results are obtained for mass $m = 6$.}
\label{fig:SMcharge2}
\end{figure}

The $q=2$ case exhibits further interesting differences. As shown in Fig.~\ref{fig:SMcharge2}(a,b), while the two-stage nature of string breaking is preserved as evident from the corresponding jumps in the total vacuum-subtracted matter density, $\langle \hat{n} \rangle - \langle \hat{n}_{\mathrm{vac}} \rangle$, and in the entanglement entropy, \(\mathcal{S}\), the intermediate plateau associated with partial string breaking is significantly reduced comparing it to the case studied in the main text in Fig.~\ref{fig:combined_charge2}(a,b). This behavior indicates an enhanced tendency toward complete string breaking when static charges $q=2$ are imposed in this configuration. During this first string-breaking event, we observe that the positron is distributed asymmetrically over the four even lattice sites surrounding the odd site with the $-q$ static charge, while the electron is asymmetrically distributed over the four odd lattice sites surrounding the even site with the $+q$ static charge. We attribute the observed instability towards full string breaking to this asymmetry in the partially broken string configuration, as illustrated in Fig.~\ref{fig:SMcharge2}(d). This feature is absent in the corresponding case discussed in the main text (see Fig.~\ref{fig:combined_charge2}(d)). Upon full string breaking, the additional positron produced is again distributed over the four even lattice sites surrounding the $-q$ static charge, while the additional electron again occupies the four odd lattice sites surrounding the $+q$ static charge, thereby achieving complete screening, as shown in Fig.~\ref{fig:SMcharge2}(e). Another notable difference relative to the scenario discussed in the main text is that, while increasing the magnetic coupling $g_B$ again shifts the transition point toward larger values, the observed trend suggests that the magnitude of this shift diminishes as $g_B$ increases as evident from Fig.~\ref{fig:SMcharge2}(a,b).

\section{Circuit construction for the minimal coupling term}\label{Sec:Circs_coupling}
Our circuit decomposition for the minimal coupling term is justified by the following argument. Here, we adopt the qudit labeling convention introduced in the main text. We make use of the identity \begin{equation}
e^{-\imath  \theta \hat{\sigma}^{x; a,b}_{\mathbf{j, e_{\mu}}}} = \hat{H}^{a,b}_{\mathbf{j, e_{\mu}}} \, e^{-\imath  \theta \hat{\sigma}^{z;a,b}_{\mathbf{j, e_{\mu}}}}\, \hat{H}^{a,b}_{\mathbf{j, e_{\mu}}},
\end{equation}
where $\{\ket{a}, \ket{b}\}$ denotes the target subspace, and $\hat{H}^{a,b}_{\mathbf{j, e_{\mu}}}$ is the Hadamard operator acting within that subspace. We further employ the relations
\begin{equation}
\begin{aligned}
(\hat{P}^{0}_{\mathbf{j}} \hat{\sigma}^{x;12}_{\mathbf{j, j+e_{\mu}}} \hat{P}^{0}_{\mathbf{j, e_{\mu}}})^2 & = \hat{{I}}_{\hat{P}^0_{\mathbf{j}}} \hat{{I}}^{1,2}_{\mathbf{j, j+e_{\mu}}} \hat{{I}}_{\hat{P}^0_{\mathbf{j, e_{\mu}}}} =  \hat{{I}}',\\
(\hat{P}^{1}_{\mathbf{j}} \hat{\sigma}^{x;01}_{\mathbf{j, j+e_{\mu}}} \hat{P}^{1}_{\mathbf{j, e_{\mu}}})^2 & = \hat{{I}}_{\hat{P}^1_{\mathbf{j}}} \hat{{I}}^{0, 1}_{\mathbf{j, j+e_{\mu}}}\hat{{I}}_{\hat{P}^1_{\mathbf{j, e_{\mu}}}} = \hat{{I}}'',
\end{aligned}
\end{equation}
where each $\hat{I}$ represents the identity operator restricted to the relevant subspace defined by either the projectors, or $\hat{\sigma}^{x;ab}_{\mathbf{j, e_{\mu}}}$ acting on the relevant qudits, 
to get 
\begin{equation}\label{unitary_coupling_s=1}
\begin{aligned}
\hat{U}_{C_{\mathbf{j, e_{\mu}}}}^{\rm MIO}(\theta) = & \sum_{n=0}^\infty\frac{\left(-\imath \theta \hat{P}^{0}_{\mathbf{j}} \hat{\sigma}^{x;12}_{\mathbf{j, j+e_{\mu}}} \hat{P}^{0}_{\mathbf{j, e_{\mu}}} + \hat{P}^{1}_{\mathbf{j}} \hat{\sigma}^{x;01}_{\mathbf{j, j+e_{\mu}}} \hat{P}^{1}_{\mathbf{j, e_{\mu}}}\right)^n}{n!}\\
= & \, \mathbb{1} - \hat{\mathrm{I}}' + \hat{P}^0_{\mathbf{j}}\, \hat{H}^{12}_{\mathbf{j, e_{\mu}}} \, e^{-\imath \theta \hat{\sigma}^{z;12}_{\mathbf{j, e_{\mu}}}}\, \hat{H}^{12}_{\mathbf{j, e_{\mu}}} \, \hat{P}^0_{\mathbf{j+e_{\mu}}}\\
& + \hat{P}^1_{\mathbf{j}}\, \hat{H}^{01}_{\mathbf{j, e_{\mu}}} \, e^{-\imath \theta \hat{\sigma}^{z;01}_{\mathbf{j, e_{\mu}}}}\, \hat{H}^{01}_{\mathbf{j, e_{\mu}}} \, \hat{P}^1_{\mathbf{j+e_{\mu}}}.
\end{aligned}
\end{equation}
Consequently, we have $\hat{\mathrm{I}}' = \hat{{I}}' + \hat{{I}}''$ denoting the identity operator on the combined subspace.

Equation~\eqref{unitary_coupling_s=1} admits a clear operational interpretation. For any seven-qudit state that fails to satisfy the relevant projector conditions, the corresponding circuit acts trivially as the identity operator. In contrast, when the projector conditions are satisfied by both left and right neighbours, the circuit applies the appropriate flip operation within the corresponding two-level subspace on the target qudit. This behavior is enforced in our construction through the use of mapping tables that encode the projector constraints, and correctly map each three-qudit state that satisfies the same kind of projector condition such that the third qudit is sent to either $\ket{3}$ or $\ket{4}$, depending on the projector condition satisfied. Conversely, any three-qudit state that does not satisfy any projector condition is mapped to a state of the third qudit distinct from $\ket{3}$ and $\ket{4}$.

\begin{table}[H]
    \centering
    \begin{tabular}{|c|c|}
        \hline
        \textbf{3-link state} & \textbf{3-qudit state} \\  
        \hline
        $\ket{110}$ & $\ket{104}$ \\  
        \hline
        $\ket{101}$ & $\ket{114}$ \\  
        \hline
        $\ket{011}$ & $\ket{014}$ \\  
        \hline
        $\ket{002}$ & $\ket{024}$ \\  
        \hline
        $\ket{020}$ & $\ket{004}$ \\  
        \hline
        $\ket{200}$ & $\ket{204}$ \\  
        \hline  
    \end{tabular}
    \caption{Mapping table for states satisfying $ \hat{P}_{\mathbf{j}}^{-1}$ condition}
    \label{tab:s_1_p-1}
\end{table}

\begin{table}[H]
    \centering
    \begin{tabular}{|c|c|}
        \hline
        \textbf{3-link state} & \textbf{3-qudit state} \\  
        \hline
        $\ket{111}$ & $\ket{103}$ \\  
        \hline
        $\ket{120}$ & $\ket{123}$ \\  
        \hline
        $\ket{102}$ & $\ket{113}$ \\  
        \hline
        $\ket{021}$ & $\ket{003}$ \\  
        \hline
        $\ket{012}$ & $\ket{013}$ \\  
        \hline
        $\ket{210}$ & $\ket{233}$ \\  
        \hline
        $\ket{201}$ & $\ket{203}$ \\  
        \hline
        
    \end{tabular}
    \caption{Mapping table for states satisfying $ \hat{P}_{\mathbf{j}}^{0}$ condition}
    \label{tab:s_1_p0}
\end{table}

To construct the complete circuit for the minimal coupling term in the spin-$1$ QLM, we begin by applying the projector verification subcircuit to the left neighbors. These subcircuits are constructed based on the corresponding mapping tables for projectors $\hat{P}^{-1}_{\mathbf{j}}$ and $\hat{P}^{0}_{\mathbf{j}}$ given in Table \ref{tab:s_1_p-1} and \ref{tab:s_1_p0}, respectively. Subsequently, the same projector verification subcircuit is applied to the right neighbors. Then, a doubly-controlled $\hat{\sigma}^{x;ab}$ rotation is performed within the appropriate subspace on the target qudit, with the controls provided by the third qudit of both sets of neighbouring links. Finally, the sequence of operations is reversed to restore all qudits to the $\{\ket{0},\ket{1},\ket{2}\}$ subspace. 

\section{Circuit construction for the plaquette term}\label{Sec:Circs_plaquette}

The plaquette term $\hat{H}_{\square}$ in \eqref{h_qudit_qlm} can be decomposed into $\hat{S}^{x}$ and $\hat{S}^{y}$ operators as
\begin{equation}\label{plaquette_term}
\begin{aligned}
    \hat{H}_{\square} &= \hat{S}_{\mathbf{j},1}^{+} \hat{S}_{\mathbf{j},2}^{-} \hat{S}_{\mathbf{j},3}^{+} \hat{S}_{\mathbf{j},4}^{-} + \text{H.c.}\\
     & = 2\Big(
\hat{S}^{x}_{\mathbf{j},1}\hat{S}^{x}_{\mathbf{j},2}\hat{S}^{x}_{\mathbf{j},3}\hat{S}^{x}_{\mathbf{j},4}
+ \hat{S}^{y}_{\mathbf{j},1}\hat{S}^{y}_{\mathbf{j},2}\hat{S}^{y}_{\mathbf{j},3}\hat{S}^{y}_{\mathbf{j},4}
\\
&\quad
+ \hat{S}^{x}_{\mathbf{j},1}\hat{S}^{y}_{\mathbf{j},2}\hat{S}^{y}_{\mathbf{j},3}\hat{S}^{x}_{\mathbf{j},4}
+ \hat{S}^{y}_{\mathbf{j},1}\hat{S}^{x}_{\mathbf{j},2}\hat{S}^{x}_{\mathbf{j},3}\hat{S}^{y}_{\mathbf{j},4}
\\
&\quad
- \hat{S}^{x}_{\mathbf{j},1}\hat{S}^{y}_{\mathbf{j},2}\hat{S}^{x}_{\mathbf{j},3}\hat{S}^{y}_{\mathbf{j},4}
- \hat{S}^{y}_{\mathbf{j},1}\hat{S}^{x}_{\mathbf{j},2}\hat{S}^{y}_{\mathbf{j},3}\hat{S}^{x}_{\mathbf{j},4}
\\
&\quad
+ \hat{S}^{x}_{\mathbf{j},1}\hat{S}^{x}_{\mathbf{j},2}\hat{S}^{y}_{\mathbf{j},3}\hat{S}^{y}_{\mathbf{j},4}
+ \hat{S}^{y}_{\mathbf{j},1}\hat{S}^{y}_{\mathbf{j},2}\hat{S}^{x}_{\mathbf{j},3}\hat{S}^{x}_{\mathbf{j},4}
\Big).
\end{aligned}
\end{equation}
where we have defined $(\mathbf{j}, 1) \equiv (\mathbf{j}, \mathbf{e}_x)$, $(\mathbf{j}, 2) \equiv (\mathbf{j + \mathbf{e}_x}, \mathbf{e}_y)$, $(\mathbf{j}, 3) \equiv (\mathbf{j + \mathbf{e}_y}, \mathbf{e}_x)$, and $(\mathbf{j}, 4) \equiv (\mathbf{j}, \mathbf{e}_y)$. All terms in $\hat{H}_{\square}$ commute with each other. This mutual commutativity arises because each term is a product of four spin operators—either $\hat{S}^{x}$ or $\hat{S}^{y}$—and any two terms differ by swapping $\hat{S}^{x} \leftrightarrow \hat{S}^{y}$ on an even number of sites.

The spin operators on different sites trivially commute, so the commutator between two such plaquette terms depends only on the number of sites where the operators differ and do not commute (i.e., $ [\hat{S}^{x}, \hat{S}^{y}] \neq 0 $). Since each pair of terms differs on an even number of such sites, the total commutator vanishes due to antisymmetry of the commutator and the even number of sign flips. Thus, all eight terms in Eq.~\eqref{plaquette_term} commute pairwise, and the evolution operator $\hat{U}_{\square}^{\rm MIO}(\theta)$ can be written as 
\begin{equation}\label{plaq_decomp_1}
\begin{aligned}
\hat{U}_{\square}^{\rm MIO}(\theta)
= {} &
e^{-\imath 2\theta \,
\hat{S}^{x}_{\mathbf{j},1}
\hat{S}^{x}_{\mathbf{j},2}
\hat{S}^{x}_{\mathbf{j},3}
\hat{S}^{x}_{\mathbf{j},4}}
\;
e^{-\imath 2\theta \,
\hat{S}^{y}_{\mathbf{j},1}
\hat{S}^{y}_{\mathbf{j},2}
\hat{S}^{y}_{\mathbf{j},3}
\hat{S}^{y}_{\mathbf{j},4}}
\\
&\times
e^{-\imath 2\theta \,
\hat{S}^{x}_{\mathbf{j},1}
\hat{S}^{y}_{\mathbf{j},2}
\hat{S}^{y}_{\mathbf{j},3}
\hat{S}^{x}_{\mathbf{j},4}}
\;
e^{-\imath 2\theta \,
\hat{S}^{y}_{\mathbf{j},1}
\hat{S}^{x}_{\mathbf{j},2}
\hat{S}^{x}_{\mathbf{j},3}
\hat{S}^{y}_{\mathbf{j},4}}
\\
&\times
e^{+\imath 2\theta \,
\hat{S}^{x}_{\mathbf{j},1}
\hat{S}^{y}_{\mathbf{j},2}
\hat{S}^{x}_{\mathbf{j},3}
\hat{S}^{y}_{\mathbf{j},4}}
\;
e^{+\imath 2\theta \,
\hat{S}^{y}_{\mathbf{j},1}
\hat{S}^{x}_{\mathbf{j},2}
\hat{S}^{y}_{\mathbf{j},3}
\hat{S}^{x}_{\mathbf{j},4}}
\\
&\times
e^{-\imath 2\theta \,
\hat{S}^{x}_{\mathbf{j},1}
\hat{S}^{x}_{\mathbf{j},2}
\hat{S}^{y}_{\mathbf{j},3}
\hat{S}^{y}_{\mathbf{j},4}}
\;
e^{-\imath 2\theta \,
\hat{S}^{y}_{\mathbf{j},1}
\hat{S}^{y}_{\mathbf{j},2}
\hat{S}^{x}_{\mathbf{j},3}
\hat{S}^{x}_{\mathbf{j},4}} .
\end{aligned}
\end{equation}
Further using $\hat{S}^{x} = \hat{\sigma}^{x;01} + \hat{\sigma}^{x;12}$ and $\hat{S}^{y} = \hat{\sigma}^{y;01} + \hat{\sigma}^{y;12}$, we can decompose each exponential term in $\hat{U}_{\square}^{\rm MIO}(\theta)$ in terms of operations on qudit subspaces. Since $[\hat{\sigma}^{i, ab}, \hat{\sigma}^{i, bc}] \neq 0$, this decomposition introduces finite digitization errors arising from the non-commutativity of the corresponding exponential terms. For example,
\begin{equation}\label{plaq_decomp_2}
\begin{aligned}
e^{-\imath 2\theta \,
\hat{S}^{x}_{\mathbf{j},1}
\hat{S}^{x}_{\mathbf{j},2}
\hat{S}^{x}_{\mathbf{j},3}
\hat{S}^{x}_{\mathbf{j},4}}
\;\approx\;
&
e^{-\imath 2\theta \,
\hat{\sigma}^{x;01}_{\mathbf{j},1}
\hat{\sigma}^{x;01}_{\mathbf{j},2}
\hat{\sigma}^{x;01}_{\mathbf{j},3}
\hat{\sigma}^{x;01}_{\mathbf{j},4}}
\\
& \underbrace{
\times
e^{-\imath 2\theta \,
\hat{\sigma}^{x;01}_{\mathbf{j},1}
\hat{\sigma}^{x;01}_{\mathbf{j},2}
\hat{\sigma}^{x;01}_{\mathbf{j},3}
\hat{\sigma}^{x;12}_{\mathbf{j},4}}
\times \cdots
}_{15\ \text{terms}}
\end{aligned}
\end{equation}

can be decomposed into 16 distinct four-body exponential terms upto some Trotter errors governed by the non-commutativity $[\hat{\sigma}^{x, 01}, \hat{\sigma}^{x, 12}] \neq 0$.

As a representative example, we present the circuit construction for $e^{-\imath \theta \, \hat{\sigma}^{x;01}_{\mathbf{j},1} \hat{\sigma}^{x;01}_{\mathbf{j},2} \hat{\sigma}^{x;01}_{\mathbf{j},3} \hat{\sigma}^{x;01}_{\mathbf{j},4}}$ here, since the circuits for all remaining terms appearing in the Trotter decomposition \ref{plaq_decomp_2} can be generated from this fundamental building block. As this operator entangles the four qudits together through the simultaneous action of $\hat{\sigma}^{x,01}$ operators, one can decompose it into Mølmer–Sørensen (MS) gates, defined as
\begin{equation}\label{eq:MS}
\operatorname{MS}^{a, b}(\alpha, \beta)=\exp \left(-\frac{\imath  \alpha}{4}\left[\sigma^{_\beta; ab} \otimes \mathbb{1}+\mathbb{1} \otimes \sigma^{_\beta; ab}\right]^2\right),
\end{equation}
where $\sigma^{_\beta; ab} = (\cos  \beta \,\hat{\sigma}^{x;ab}\pm \sin \beta \, \hat{\sigma}^{y;ab})$. Firstly, we apply MS gates between qudit pairs ($A$, $D$), ($B$, $D$), ($C$, $D$) which entangles them together. Subsequently, $\operatorname{Rz}^{ab}(\pi - 2\theta)$ is applied to qudit $A$ in the $\{\ket{0}, \ket{1}\}$ subspace. This gate can be implemented as $\operatorname{Rz}^{ab}(\phi) = \operatorname{VRz}^{a}(\phi) \operatorname{VRz}^{b}(\phi)$. The entanglement ensures that this gate has the desired effect on the four-qudit state. Then, the MS gates are applied in reverse to disentangle the qudits. The two $\operatorname{Rz}^{ab}(\pi/2)$ gates on qudit $D$ in the beginning and the end are used to ensure that the rotation by $\theta$ is applied overall. 

An alternate way to decompose the operator $e^{-\imath \theta \, \hat{\sigma}^{x;01}_{\mathbf{j},1} \hat{\sigma}^{x;01}_{\mathbf{j},2} \hat{\sigma}^{x;01}_{\mathbf{j},3} \hat{\sigma}^{x;01}_{\mathbf{j},4}}$, inspired by circuit constructions for analogous three-body interactions using tunable MS gates \cite{Andrade_2022, Meth_Zhang_Haase_Edmunds_Postler_Jena_Steiner_Dellantonio_Blatt_Zoller_et}, is given as
\begin{equation}\label{plaq_decomp_3}
\begin{aligned}
e^{-\imath \theta \, \hat{\sigma}^{x;01}_{\mathbf{j},1} \hat{\sigma}^{x;01}_{\mathbf{j},2} \hat{\sigma}^{x;01}_{\mathbf{j},3} \hat{\sigma}^{x;01}_{\mathbf{j},4}}
= & \; e^{-\imath \frac{\pi}{4} \hat{\sigma}^{x;01}_{\mathbf{j},1} \, \hat{\sigma}^{y;01}_{\mathbf{j},2}} e^{-\imath \frac{\pi}{4} \hat{\sigma}^{z;01}_{\mathbf{j},2} \, \hat{\sigma}^{y;01}_{\mathbf{j},3}} \\
& \times e^{-\imath \theta \hat{\sigma}^{z;01}_{\mathbf{j},3} \, \hat{\sigma}^{x;01}_{\mathbf{j},4}} \\ 
& \times e^{\imath \frac{\pi}{4} \hat{\sigma}^{z;01}_{\mathbf{j},2} \, \hat{\sigma}^{y;01}_{\mathbf{j},3}} e^{\imath \frac{\pi}{4} \hat{\sigma}^{x;01}_{\mathbf{j},1} \, \hat{\sigma}^{y;01}_{\mathbf{j},2}} .
\end{aligned}
\end{equation}

\bibliography{biblio}

\end{document}